\documentclass[authoryear,final,3p,times]{elsarticle}
\usepackage{amssymb}
\usepackage{amsthm}
\usepackage{amsmath}
\usepackage{bm}
\usepackage{array}
\usepackage{ulem}
\usepackage{longtable}
\usepackage{CJK}
\usepackage{setspace}
\usepackage{multirow}
\usepackage{siunitx}
\usepackage{threeparttable}
\usepackage{verbatim}
\usepackage{rotating}
\usepackage{xcolor}
\usepackage{booktabs}
\usepackage[justification=centering]{caption}
\usepackage{lineno}
\usepackage{setspace}
\usepackage{lscape}
\usepackage{makecell}
\usepackage{threeparttable}
\usepackage{float}
\usepackage{graphicx}
\usepackage{lineno}
\allowdisplaybreaks[4]
\usepackage[utf8]{inputenc}
\usepackage{textcomp}
\DeclareUnicodeCharacter{2217}{\ensuremath{\ast}}

\begin{document}

\begin{spacing}{1.0}
\begin{frontmatter}

\journal{  }

\title{Physics-informed neural network for inverse modeling of granular flows}


\author[label1]{Bing Wan}

\author[label1,label2]{Bidan Zhao\corref{cor1}}
\cortext[cor1]{Corresponding author}
\ead{bdzhao@cup.edu.cn}

\author[label1,label3]{Junwu Wang\corref{cor2}}
\cortext[cor2]{Corresponding author}
\ead{jwwang@cup.edu.cn}

\address[label1]{College of New Energy Science and Engineering, China University of Petroleum-Beijing, Beijing 102249, China}

\address[label2]{State Key Laboratory of Heavy Oil Processing, China University of Petroleum-Beijing, Beijing 102249, China}

\address[label3]{State Key Laboratory of Deep Geothermal Resources, China University of Petroleum-Beijing, Beijing 102249, China}

\begin{abstract}
Granular flows are ubiquitous in natural and industrial systems, yet their complex dynamics remain difficult to characterize. For inverse problems involving unknown inlet, outlet, and wall boundary conditions, where CFD simulations are challenging, reconstructing complete flow fields from sparse observations constitutes a challenging inverse problem. In this study, a physics-informed neural network framework driven by both physical mechanisms and measurement data is developed to reconstruct the steady-state full-field distribution of granular flows in a pipe. The proposed approach integrates sparse measurement data with governing equations and constitutive relations and is trained using high-fidelity datasets generated by CFD solutions of a continuum model. The framework incorporates a dimensionless loss formulation, physics-informed initialization, dynamic global weighting, and a locally weighted granular temperature data-loss strategy. These treatments enable accurate reconstruction of the complete flow-field evolution. This work establishes a robust methodological framework for flow-field reconstruction in complex granular flow systems.
\end{abstract}

\begin{keyword}
machine learning; PINN; granular flow; flow-field reconstruction; inverse modeling
\end{keyword}
\end{frontmatter}

\section{Introduction}
Granular matter is a complex system composed of numerous discrete particles interacting through contact forces, whose individual motions follow Newton's laws of motion \citep{ge2019multiscale}. When the internal stress state of a granular assembly changes or external forces are applied, the particle ensemble may exhibit fluid-like motion. This dynamic state is referred to as granular flow. Granular flows are ubiquitous in nature, daily life, and industrial processes, including chemical processing, metallurgical operations, food manufacturing, and mining \citep{henann2013predictive,henann2014continuum,jerolmack2019viewing}. Therefore, understanding the complex dynamic evolution of granular flows is essential for the design, optimization, and scale-up of particulate systems \citep{goldhirsch2003rapid,wang2020continuum}.

Current studies of granular flow mainly rely on numerical simulations and experimental measurements. Among numerical approaches, Eulerian models \citep{wang2020continuum} treat the particulate phase as a continuum and describe its macroscopic behavior using conservation equations, whereas the Discrete Element Method (DEM) \citep{zhao2020cfd} explicitly resolves particle-particle interactions and provides particle-scale trajectories and positions. However, numerical simulations are computationally expensive, especially for large-scale systems, and more importantly, they are impossible to apply directly to inverse problems with unknown inlet and outlet conditions.
Experimental measurements provide an alternative means of obtaining flow-field information. Particle Image Velocimetry (PIV) \citep{wang2024effects,wang2018scale,zhang2023configuration,zhu2023particle} combines high-speed imaging with image-correlation algorithms and can rapidly obtain quasi-continuous two-dimensional velocity distributions. However, limited temporal and spatial resolutions and the opacity of dense granular systems restrict its applicability in three-dimensional (3D),high-fidelity full-field measurements. Capacitance probes \citep{acree1989quantitative,hage1997guarded,richtberg2005characterization} measure local solid volume fraction and its fluctuations through variations in capacitance signals near the probe. With dual-probe or multi-probe configurations, particle velocities can also be estimated using cross-correlation analysis. Optical fiber probes \citep{qi2022random,yasui1958characteristics,wei2020experimental,liu2020cluster} measure local solid volume fraction based on the correlation between reflected light intensity and solid volume fraction, and local particle velocity can be further estimated using time-of-flight or cross-correlation methods in dual-fiber configurations. Nevertheless, both capacitance and optical fiber probes both capacitance and optical fiber probes are intrusive measurement techniques, and measure the data from limited measurement points, they are also extremely difficult in 3D, high-fidelity full-field measurements. Electrical Capacitance Tomography (ECT) \citep{wang2006study,zhang2025combing,niedostatkiewicz2010application} can provide non-intrusive cross-sectional solid volume fraction distributions for dynamic monitoring, but its spatial resolution is relatively low and the reconstructed results are sensitive to the permittivity distribution and inverse algorithm. X-ray Computed Tomography (CT) \citep{luan6248507micro,bieberle2016combined,verma2014bubble} enables three-dimensional reconstruction of internal granular structures and can provide information such as particle position, shape, coordination number, porosity, and local solid volume fraction. However, it requires sophisticated instrumentation, incurs high experimental cost, and its accuracy is affected by noise and reconstruction errors. Although these measurement techniques have distinct advantages, long-term sensor accuracy and durability remain challenging \citep{zhu2024advanced}. More importantly, existing techniques still cannot simultaneously provide non-intrusive, high-resolution, full-field information of multiple physical quantities, which should at least  include the solid volume fraction, particle velocity and temperature in granular flows. Therefore, reconstructing full-field velocity and solid volume fraction fields from limited measurements has become a critical challenge in granular-flow measurement.

In recent years, Physics-Informed Neural Networks (PINN) \citep{raissi2019physics} have emerged as a promising machine-learning framework that integrates data-driven learning with physical constraints. By embedding governing equations, boundary conditions, and initial conditions into the loss function, PINN enforce physical consistency during training and learn mappings between input spatial-temporal coordinates and physical variables. Since the pioneering work of Raissi et al. \citep{raissi2019physics}, PINN have been widely applied to forward and inverse problems in fluid mechanics and related fields \citep{zhang2023physics,shokouhi2021physics,deng2021application,lu2021surrogate,zhang2025general}. Recent developments have focused on hyperparameter optimization \citep{jagtap2020adaptive,pu2021data,zhang2019quantifying}, sampling strategies and data processing \citep{wu2023comprehensive}, and loss-function design and regularization methods \citep{maddu2022inverse,mcclenny2023self}, which have significantly improved the capability of PINN for complex physical systems.
Several studies have extended PINN to granular systems. Wan et al. \citep{wan2025physics} developed a PINN framework for solving forward problems of granular flow in the homogeneous cooling state and used sparse data for inverse parameter identification. Zolfaghari and Jamali \citep{zolfaghari2026non} constructed a PINN framework based on the Nonlocal Granular Fluidity (NGF) model, enabling forward prediction and inverse parameter identification for dense granular flows. With limited velocity-field data, their method reconstructed pressure and stress fields and inferred nonlocal model parameters. Baldoni et al. \citep{baldoni2025rheological} proposed a PINN-based method for rheological parameter identification in granular materials. Taking granular column collapse as a benchmark case, they inferred static and dynamic friction coefficients using velocity-field and free-surface information, while also reconstructing pressure fields that are difficult to measure experimentally. Hu et al. \citep{hu2026reconstruction} proposed a GF-PINN framework for dense granular flows by embedding the Navier--Stokes equations and granular rheology into the loss function, allowing pressure-field reconstruction and material-parameter inference from velocity-field data. In addition, Chen et al. \citep{chen2021physics} embedded the Population Balance Equation (PBE) into a physics-informed deep-learning framework for modeling particle aggregation and breakage processes, enabling both forward prediction of particle number-density distributions and inverse identification of PBE parameters. Zhou et al. \citep{zhou2025physics} proposed a PINN model for multi-particle interaction forces by combining ResNet with physical constraints derived from Newton's third law, reducing the computational cost of DEM-based force evaluation. Su et al. \citep{su2024thermodynamics} developed a Thermodynamics-Informed Neural Network (TINN) for elastoplastic constitutive modeling of granular materials by explicitly incorporating thermodynamic constraints into the network. Despite these advances, PINN for granular flows remains at an early stage because granular flows involve strong nonlinearity, multiscale coupling, and collision-induced nonequilibrium effects. Strong local gradients, particle clustering, and stiffness in the governing equations may lead to gradient imbalance during PINN training \citep{cuomo2022scientific}. Moreover, in practical experimental conditions, measured flow-field data are usually sparse, and complete inlet/outlet- wall- boundary conditions are often unavailable, which further increases the difficulty of inverse reconstruction.

The objective of this study is to investigate the feasibility of reconstructing the full flow field of granular flow from sparse data using a PINN framework, thereby providing a new paradigm for full-field information acquisition in granular flow systems. Unlike existing granular-flow PINN studies that often assume constant solids viscosity, the present work employs the kinetic theory of granular flow (KTGF) \citep{gidaspow1994multiphase} to close the solid-phase stress tensor, so that transport coefficients such as viscosity and solids thermal conductivity vary spatially and temporally with local flow conditions. Three strategies are incorporated into the proposed PINN framework: loss-function nondimensionalization, physics-informed initialization based on measured samples, and local weighting of granular-temperature data loss according to physical characteristics. The remainder of this paper is organized as follows. Section 2 introduces the governing equations and constitutive relations for granular pip flow and simulates this system for getting sample data and reference data for PINN. Section 3 presents the proposed PINN framework and implementation strategies. Section 4 demonstrates the reconstruction of the full dynamic evolution of granular flow from limited sparse data and investigates the influence of sparse-data sampling strategies on reconstruction accuracy. Finally, Section 5 summarizes the main conclusions.

\section{Governing equations for granular flow}
In this study, a standard Eulerian-Eulerian model was used to simulate particle flow in a pipe, and the particle phase was treated as a continuum \citep{wang2020continuum,gidaspow1994multiphase}. Within the continuum-mechanics framework, the Navier-Stokes equations combined with kinetic theory of granular flow describe the mass and momentum conservation equations as follows:
\begin{equation}
      \frac{\partial(\varepsilon_s \rho_s)}{\partial t} + \nabla \cdot \left(\varepsilon_s \rho_s \mathbf{u}_s\right) = 0,
\end{equation}
\begin{equation}
      \frac{\partial(\varepsilon_s \rho_s \mathbf{u}_s)}{\partial t} + \nabla \cdot \left(\varepsilon_s \rho_s \mathbf{u}_s \mathbf{u}_s\right) = - \nabla p_s + \nabla \cdot \boldsymbol{\tau}_s,
\end{equation}
Where $\mathbf{u}_s$ and $\varepsilon_s$ denote the solid velocity and solid volume fraction, respectively. The stress tensor is assumed to be linearly related to the rate-of-strain tensor:
\begin{equation}
      \boldsymbol{\tau}_s=\left[\mu_s\left(\nabla \mathbf{u}_s+\nabla \mathbf{u}_s^T\right)+\left(\lambda_s-\frac{2}{3}\mu_s\right)(\nabla\cdot\mathbf{u}_s)\mathbf{I}\right].
\end{equation}
When the kinetic theory of granular flow (KTGF) was used to close the solid-phase stress, the granular-temperature equation must also be solved \citep{gidaspow1994multiphase}:
\begin{equation}
      \frac{3}{2}\left[\frac{\partial(\varepsilon_s\rho_s\Theta_s)}{\partial t}+\nabla\cdot(\varepsilon_s\rho_s\mathbf{u}_s\Theta_s)\right]=(-p_s\mathbf{I}+\boldsymbol{\tau}_s):\nabla\mathbf{u}_s-\nabla\cdot\mathbf{q}_s-\gamma.
\end{equation}
where $\Theta_s$ denote the granular temperature. Following Lun et al. \citep{lun1984kinetic}, the solid-phase pressure and solids viscosity are calculated as follows: The solid-phase pressure consists of a kinetic contribution $p_{s,kin}$ and a collisional contribution $p_{s,col}$:
\begin{equation}
      p_s=p_{s,kin}+p_{s,col},
\end{equation}
where
\begin{equation}
      p_{s,kin}=\varepsilon_s\rho_s\Theta_s,
\end{equation}
\begin{equation}
      p_{s,col}=2\rho_s(1+e)\varepsilon_s^2 g_0\Theta_s.
\end{equation}
The solid shear viscosity $\mu_s$ and bulk viscosity $\lambda_s$ are expressed as:
\begin{equation}
\mu_s=\frac{4}{5}\varepsilon_s^2\rho_s d_p g_0(1+e)\sqrt{\frac{\Theta_s}{\pi}}+\frac{10\rho_s d_p\sqrt{\pi\Theta_s}}{96(1+e)g_0}\left[1+\frac{4}{5}(1+e)\varepsilon_s g_0\right]^2,
\label{eq:particle_viscosity}
\end{equation}
\begin{equation}
\lambda_s=\frac{4}{3}\varepsilon_s^2\rho_s d_p g_0(1+e)\sqrt{\frac{\Theta_s}{\pi}}.
\end{equation}
The energy equation is closed using the Fourier's law for conductive transport:
\begin{equation}
\mathbf{q}_s=-k_s\nabla\Theta_s,
\end{equation}
where the solids thermal conductivity is:
\begin{equation}
k_s=\frac{150\rho_s d_p\sqrt{\Theta_s\pi}}{384(1+e)g_0}\left[1+\frac{6}{5}\varepsilon_s g_0(1+e)\right]^2+2\rho_s d_p\varepsilon_s^2 g_0(1+e)\sqrt{\frac{\Theta_s}{\pi}}.
\label{eq:solids_thermal_conductivity}
\end{equation}
The collisional dissipation of granular energy is expressed as:
\begin{equation}
\gamma=\frac{12\left(1-e^2\right)g_0}{d_p\sqrt\pi}\rho_s\varepsilon_s^2\Theta_s^{\frac{3}{2}},
\label{eq:collisional_dissipation}
\end{equation}
where $g_0$ is the radial distribution function, which is defined as:
\begin{equation}
g_0=\left[1-\left(\frac{\varepsilon_s}{\varepsilon_{s,max}}\right)^\frac{1}{3}\right]^{-1}.
\end{equation}
For monodisperse rigid spherical particles, $\varepsilon_{s,max}=0.63$.
The numerical parameters used in the CFD simulations are listed in Table \ref{CFD simulation}.

\begin{table}[htbp]
    \centering
    \caption{CFD simulation parameters for granular pipe flow} \label{CFD simulation}
    \begin{tabular}{cc}
\hline
Parameter & Value \\
\hline
Particle diameter, $d_p\,(m)$ & $1.2\times10^{-3}$ \\
Particle density, $\rho_s\,(kg/m^3)$ & $2000$ \\
Inlet solid volume fraction, $\varepsilon_0$ & $0.1$ \\
Inlet solid velocity, $u_0\,(m/s)$ & $2$ \\
Inlet granular temperature, $\Theta_0\,(m^2/s^2)$ & $1.0\times10^{-4}$ \\
Side width of simulation domain, $L\,(m)$ & $0.12$ \\
Side height of simulation domain, $H\,(m)$ & $0.84$ \\
Particle-particle coefficient of restitution, $e$ & $0.95$ \\
Particle-wall coefficient of restitution, $e$ & $0.95$ \\
Total number of grid cells & $80\times560$ \\
\hline
\end{tabular}
\end{table}

\begin{figure}[htbp]
    \centerline{
        \includegraphics[
            width=1.0\textwidth,
            trim=0.5cm 3.0cm 0.5cm 3.0cm,
            clip
        ]{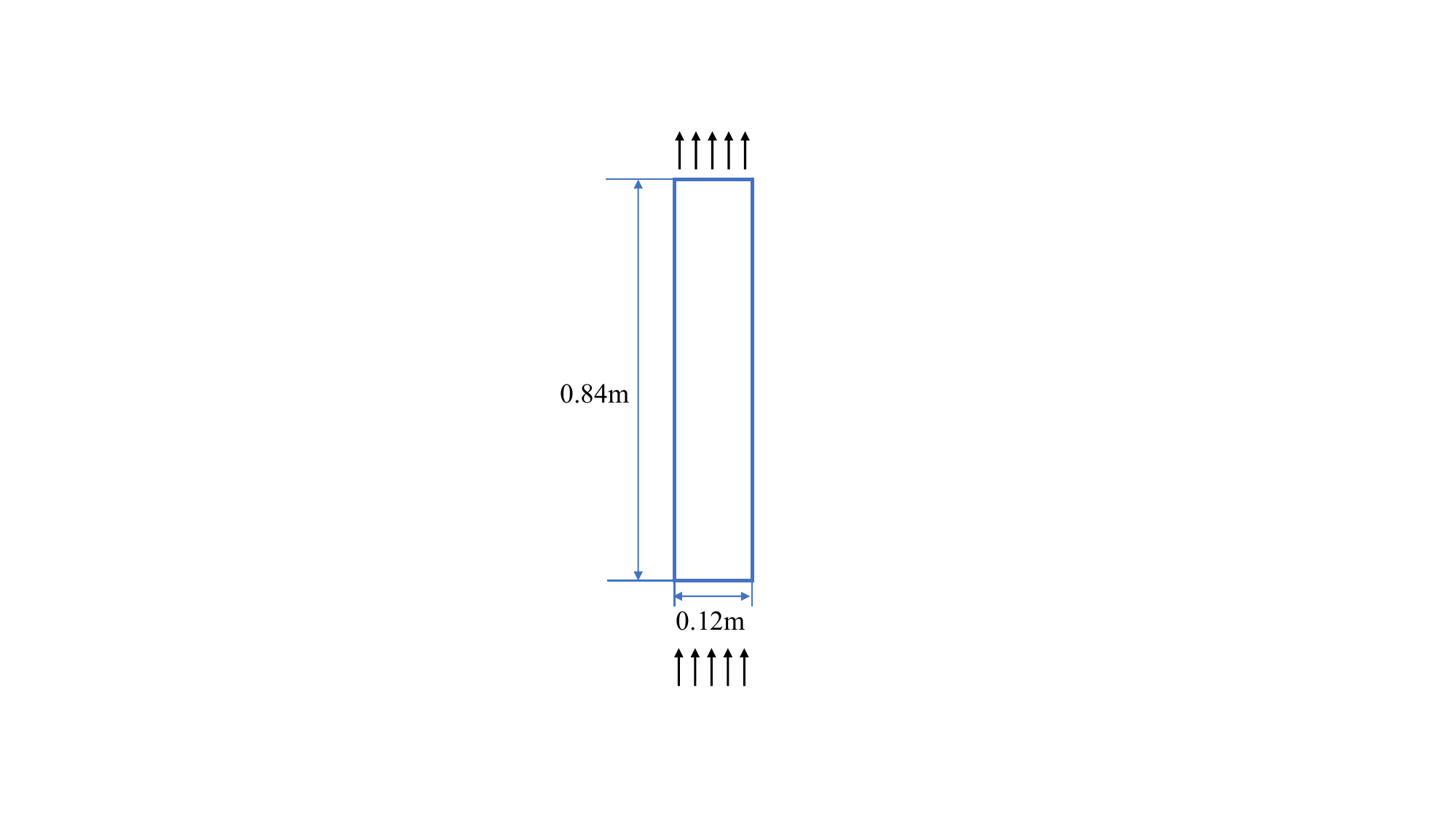}
    }
    \caption{The sketch of the granular pip flow}
    \label{1_CFD_simulation_domain}
\end{figure}
As shown in Figure~\ref{1_CFD_simulation_domain}, the left and right boundaries of the pipe are assumed to be no-slip, adiabatic walls. The bottom boundary is the particle inlet, where the inlet granular temperature is 0.0001 $\rm m^2 / \rm s^2$, the inlet solid velocity is 2 m/s, and the inlet solid volume fraction is 0.1. The top boundary is a free outflow for granular flow. The midpoint of the pipe base is defined as the coordinate origin (0, 0). Particle gravity is neglected to isolate the effects of particle-particle interactions and flow confinement on the velocity, solid volume fraction, and granular temperature fields. The reference data used as data supervision points in the PINN loss function were obtained from time-averaged numerical-simulation results over 5-10 s.

\section{Physics-informed neural network}
\begin{figure}[htbp]
    \centerline{
        \includegraphics[
            width=1.0\textwidth,
            trim=0.5cm 0.5cm 0.5cm 0.5cm,
            clip
        ]{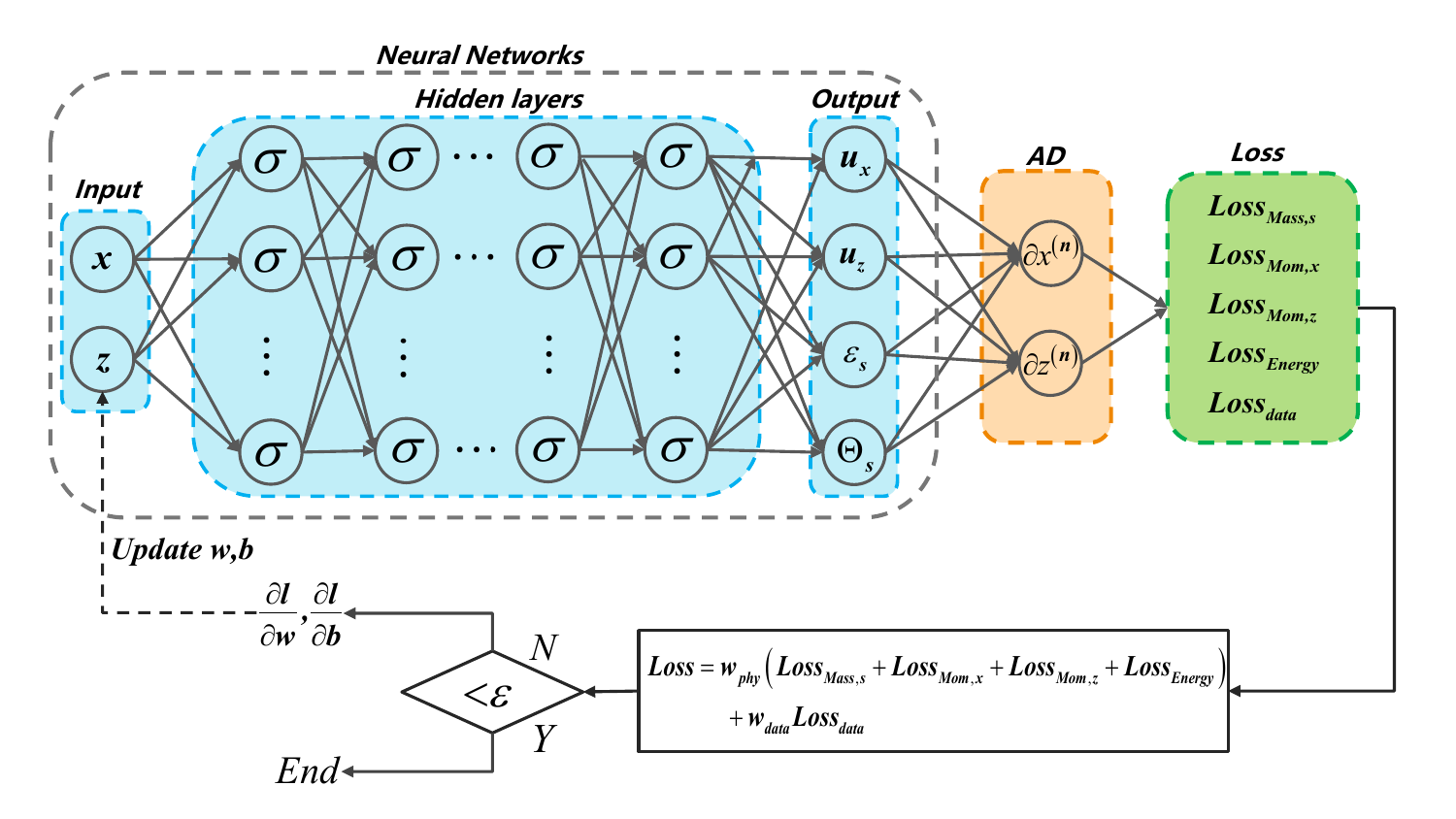}
    }
    \caption{PINN framework for reconstructing granular flow}
    \label{1_PINN_framework}
\end{figure}

In the numerical simulations, the primary quantities of interest are the spatial distributions of solid velocity, solid volume fraction, and granular temperature after the granular pipe flow reaches a steady state under the specified inlet and boundary conditions. The objective of this study is therefore to reconstruct the steady-state solid velocity field, solid volume fraction field, and granular temperature field from sparse flow-field measurement data. The PINN solves the steady-state governing equations, including the mass, momentum, and granular temperature equations, with all time-derivative terms set to zero. Figure \ref{1_PINN_framework} illustrates the PINN framework for reconstructing granular pipe flow. The spatial coordinates aligning with the center of CFD computational grid are used as inputs, and the solid velocity, solid volume fraction, and granular temperature are predicted as outputs.

The fully connected neural-network architecture used in this work can be written as:
\begin{equation}
\bm{a}^0 = (x, z)
\end{equation}
\begin{equation}
\bm{a}^k = \sigma\left(\bm{w}^k \bm{a}^{k-1} + \bm{b}^k\right), \quad 1 \le k \le L-1
\end{equation}
\begin{equation}
\bm{a}^k = \sigma_{out}\left(\bm{w}^k \bm{a}^{k-1} + \bm{b}^k\right), \quad k = L
\end{equation}

The network contains five hidden layers, each with 60 neurons. The output of the final layer approximates the true solution and is denoted by $\hat{u}(x,z)\approx \bm{a}^L$. The symbols $\bm{w}^k$ and $\bm{b}^k$ denote the weight matrix and bias vector of the $k^{th}$ layer, respectively, and $\sigma$ denotes the nonlinear activation function. The hyperbolic tangent function (tanh) is used in the hidden layers. For the output layer, appropriate activation functions must be selected to ensure that the network outputs remain within physically admissible ranges and to preserve the stability of gradient descent. The two velocity components, $u_x$ and $u_z$, are represented by linear outputs. For the solid volume fraction, the output activation function is $\varepsilon_{s,max}\cdot sigmoid=\frac{\varepsilon_{s,max}}{1+e^{-a}}$, which satisfies the physical constraint $0\le\varepsilon_s\le\varepsilon_{s,max}$ and prevents numerical divergence in the radial distribution function $g_0$. For the granular temperature, the output activation function is $\mathrm{softplus}\left(a\right)=\ln{\left(1+e^a\right)}$, which guarantees that the granular temperature remains positive, satisfies the corresponding physical constraint, and prevents numerical divergence during the evaluation of the transport coefficients and the granular-temperature equation, thereby avoiding program failure. Xavier initialization is used for the neural-network weights and the hidden-layer biases are initialized to zero. In the output layer, the bias corresponding to the solid volume fraction is set to $-0.1$ according to its physical range, which accelerates convergence and helps avoid unfavorable local optima; after the sigmoid transformation, this gives $\varepsilon_{s,max}\cdot sigmoid\approx0.30$. This initialization keeps the initial solid volume fraction field close to the midpoint of the measured range and improves the physical plausibility of the initial prediction. Except for $\varepsilon_s$, all other output-layer biases are initialized to zero. This strategy incorporates prior information on the solid volume fraction while retaining the stochastic character of the neural-network initialization.

In PINN, solving the partial differential equation system is transformed into an optimization problem in which the network weights and biases are updated iteratively to minimize the loss function. The residual of the mass conservation equation at the physics constraint points is denoted by $e_{m,s}$. The residuals of the momentum conservation equations in the two coordinate directions are denoted by $e_{u,x}$ and $e_{u,z}$, respectively, and the residual of the energy conservation equation (or the granular temperature equation) is denoted by $e_\Theta$:
\begin{equation}
e_{m,s}=\frac{\partial\left(\varepsilon_s\rho_s u_x\right)}{\partial x}+\frac{\partial\left(\varepsilon_s\rho_su_z\right)}{\partial z},
\end{equation}
\begin{equation}
e_{u,x}=\frac{\partial\left(\varepsilon_s\rho_su_xu_x\right)}{\partial x}+\frac{\partial\left(\varepsilon_s\rho_su_xu_z\right)}{\partial z}+\frac{\partial p_s}{\partial x}+\left(\frac{\partial\tau_{s,xx}}{\partial x}+\frac{\partial\tau_{s,zx}}{\partial z}\right),
\end{equation}
\begin{equation}
e_{u,z}=\frac{\partial\left(\varepsilon_s\rho_su_zu_x\right)}{\partial x}+\frac{\partial\left(\varepsilon_s\rho_su_zu_z\right)}{\partial z}+\frac{\partial p_s}{\partial z}+\left(\frac{\partial\tau_{s,xz}}{\partial x}+\frac{\partial\tau_{s,zz}}{\partial z}\right),
\end{equation}
\begin{equation}
e_{\Theta}=\frac{3}{2}\left(\frac{\partial(\varepsilon_s\rho_s u_x\Theta_s)}{\partial x}+\frac{\partial(\varepsilon_s\rho_s u_z\Theta_s)}{\partial z}\right)-(-p_s\mathbf{I}+\boldsymbol{\tau}_s):\nabla\mathbf{u}_s+\nabla\cdot\mathbf{q}_s+\gamma.
\end{equation}
The total PINN loss is obtained by weighting the dimensionless mass-conservation loss $Loss_{Mass,s}$ at the physics constraint points, the momentum-conservation losses $Loss_{Mom,x}$ and $Loss_{Mom,z}$ in the two coordinate directions, the energy-equation loss $Loss_{Energy}$, and the data loss $Loss_{data}$ at the data supervision points:
\begin{align}
Loss=w_{phy}(Loss_{Mass,s}+Loss_{Mom,x}+Loss_{Mom,z}+Loss_{Energy})+w_{data}Loss_{data},
\end{align}
where the individual loss terms are defined as follows:
\begin{equation}
Loss_{Mass,s}=\frac{1}{N_f}\sum_{i=1}^{N_f}\left(\left\|e_{m,s}^{i}\frac{L}{\rho_s u_0}\right\|_2^2\right),
\end{equation}
\begin{equation}
Loss_{Mom,x}=\frac{1}{N_f}\sum_{i=1}^{N_f}\left(\left\|e_{u,x}^{i}\frac{L}{\rho_s u_0^2}\right\|_2^2\right),
\end{equation}
\begin{equation}
Loss_{Mom,z}=\frac{1}{N_f}\sum_{i=1}^{N_f}\left(\left\|e_{u,z}^{i}\frac{L}{\rho_s u_0^2}\right\|_2^2\right),
\end{equation}
\begin{equation}
Loss_{Energy}=\frac{1}{N_f}\sum_{i=1}^{N_f}\left(\left\|e_{\Theta}^{i}\frac{L}{\rho_s u_0^3}\right\|_2^2\right),
\end{equation}
\begin{equation}
Loss_{data}=\frac{1}{N_{data}}\sum_{i=1}^{N_{data}}\left[\left(\frac{u_x-u_x^{cfd}}{u_0}\right)^2+\left(\frac{u_z-u_z^{cfd}}{u_0}\right)^2+\left(\varepsilon_s-\varepsilon_s^{cfd}\right)^2+\lambda_{\Theta}\left(\frac{\Theta_s-\Theta_s^{cfd}}{u_0^2}\right)^2\right]
\end{equation}

This work focuses on an inverse problem, where the full flow field is reconstructed from sparse flow-field measurement data via the proposed PINN. No measured particle data are prescribed at the inlet, no gradient constraints are imposed on the free outflow boundary, and no wall boundary condition constraints are introduced. Notably, CFD simulation is unable to solve this type of inverse problem that lack inlet and outlet boundary conditions and wall boundary, but PINN can effectively incorporate limited data to obtain accurate predictions of the whole granular pip flow.

All loss functions are nondimensionalized by dividing each residual by its corresponding characteristic scale. The characteristic velocity is the inlet velocity $u^\ast=u_0=2 $ m/s, the characteristic density is the particle density $\rho^\ast=\rho_s=2000$ $\mathrm{kg/m^3}$, the characteristic length is the pipe width $L^\ast=L=0.12 $ $\mathrm{m}$.
Moreover, because the granular temperature spans multiple scales, a local weighting model $\lambda_{\Theta}$ is introduced. This model uses the inverse dimensionless temperature to adjust the local weight of the temperature loss at each data supervision point, thereby amplifying the contribution of low-temperature data to the overall temperature loss. This treatment enhances the role of data supervision in the total PINN loss during iterative optimization. The local weighting model $\lambda_{\Theta}$ in $Loss_{data}$ is written as:
\begin{equation}
\lambda_{\Theta}=\frac{u_0^2}{\Theta_s^{cfd}}
\end{equation}

\begin{table}[htbp]
\centering
\caption{Dimensions used for each sub-loss term}
\begin{tabular}{ccc}
\hline
Loss type & Sub-loss term & Dimension \\
\hline
\multirow{4}{*}{Governing-equation loss}
& Mass conservation equation & $\dfrac{\rho_s u_0}{L}$ \\
& Momentum conservation equation (x-direction) & $\dfrac{\rho_s u_0^2}{L}$ \\
& Momentum conservation equation (z-direction) & $\dfrac{\rho_s u_0^2}{L}$ \\
& Energy conservation equation & $\dfrac{\rho_s u_0^3}{L}$ \\
\hline
\multirow{4}{*}{Data loss}
& Velocity in the x-direction & $u_0$ \\
& Velocity in the z-direction & $u_0$ \\
& Solid volume fraction & $1$ \\
& Granular temperature & $u_0^2$ \\
\hline
\end{tabular}
\end{table}

The weight coefficients determine the contribution of each loss term to the total loss. In this study, a dynamic global weighting strategy is adopted \citep{li2022dynamic}, where $w_{phy}$ denotes the weight of the governing-equation loss and $w_{data}$ denotes the weight of the data loss, satisfying $w_{phy}+w_{data}=1$.
The data-loss weight $w_{data}$ is adjusted as follows:
\begin{equation}
w_{data}=w_{min}+\left(w_{max}-w_{min}\right)\frac{1}{1+\exp\left[\beta\left(\frac{k}{K}-c\right)\right]} .
\end{equation}
Here, $w_{max}$ and $w_{min}$ are the upper and lower bounds of the weighting coefficient, respectively. In this work, $w_{max}=0.7$ and $w_{min}=0.3$. $k$ is the current iteration step, $K$ is the total number of iterations, and $\beta$ and $c$ control the variation rate and inflection point of the weight, respectively. In this study, $\beta=10$ and $c=0.5$. With this strategy, the weight of the governing-equation loss gradually increases from 0.3 during the early stage of training, while the data-loss weight decreases from 0.7. At the later stage of training, the data-loss weight gradually decreases to 0.3, and the governing-equation loss weight increases to 0.7, thereby strengthening the physical constraints and improving the stability and physical consistency of the results.

In PINN research, gradient descent algorithms are employed to optimize network hyperparameters. During iterative optimization, numerous optimizers are available; those commonly used for PINN optimization include Adam \citep{kingma2014adam} and L-BFGS \citep{byrd1995limited}. The Adam optimizer, due to its ability to dynamically adjust the learning rate and help the model escape local minima, is frequently used for fine-tuning the initial solution during network training, making it suitable for a wide range of deep learning tasks. Through multiple tests, it was found that after a certain number of iterations with the Adam optimizer, the L-BFGS algorithm provided negligible improvement for this system; therefore, this paper employs only the Adam optimizer for iterative training, with an initial learning rate of 0.0001.

\section{Results and Discussion}
This work aims to reconstruct the full flow field from sparse flow-field measurement data without prescribing inlet, outlet, and wall boundary conditions, while simultaneously KTGF-derived quantities including solids viscosity, solids thermal conductivity, collisional dissipation, and solid-phase pressure. The sparse dataset is extracted from time-averaged flow fields generated via 5-10 s of Eulerian continuum simulation. Furthermore, beyond the results presented in the main body of this paper, the effects of the number of data-supervision heights and the sampling strategy for data supervision points are systematically investigated. Since these supplementary cases yield consistent trends and conclusions, their corresponding results are documented in the Supporting Information.

The accuracy of the PINN solution is evaluated using contour plots of absolute and relative errors over the entire flow field, together with relative-error profiles at selected axial heights. The absolute and relative errors are defined as follows:
\begin{equation}
\textit{Absolute error}=\hat{y}_i-y_i
\end{equation}
\begin{equation}
\textit{$L_2$ relative error}=\frac{\sqrt{\frac{1}{N}\sum_{i=1}^{N}(\hat{y}_i-y_i)^2}}{\sqrt{\frac{1}{N}\sum_{i=1}^{N}y_i^2}}
\end{equation}
where $\hat{y}$ denotes the PINN prediction and y denotes the CFD-reference value.

\begin{figure}[htbp]
      \centerline{\includegraphics[width=0.75\textwidth]{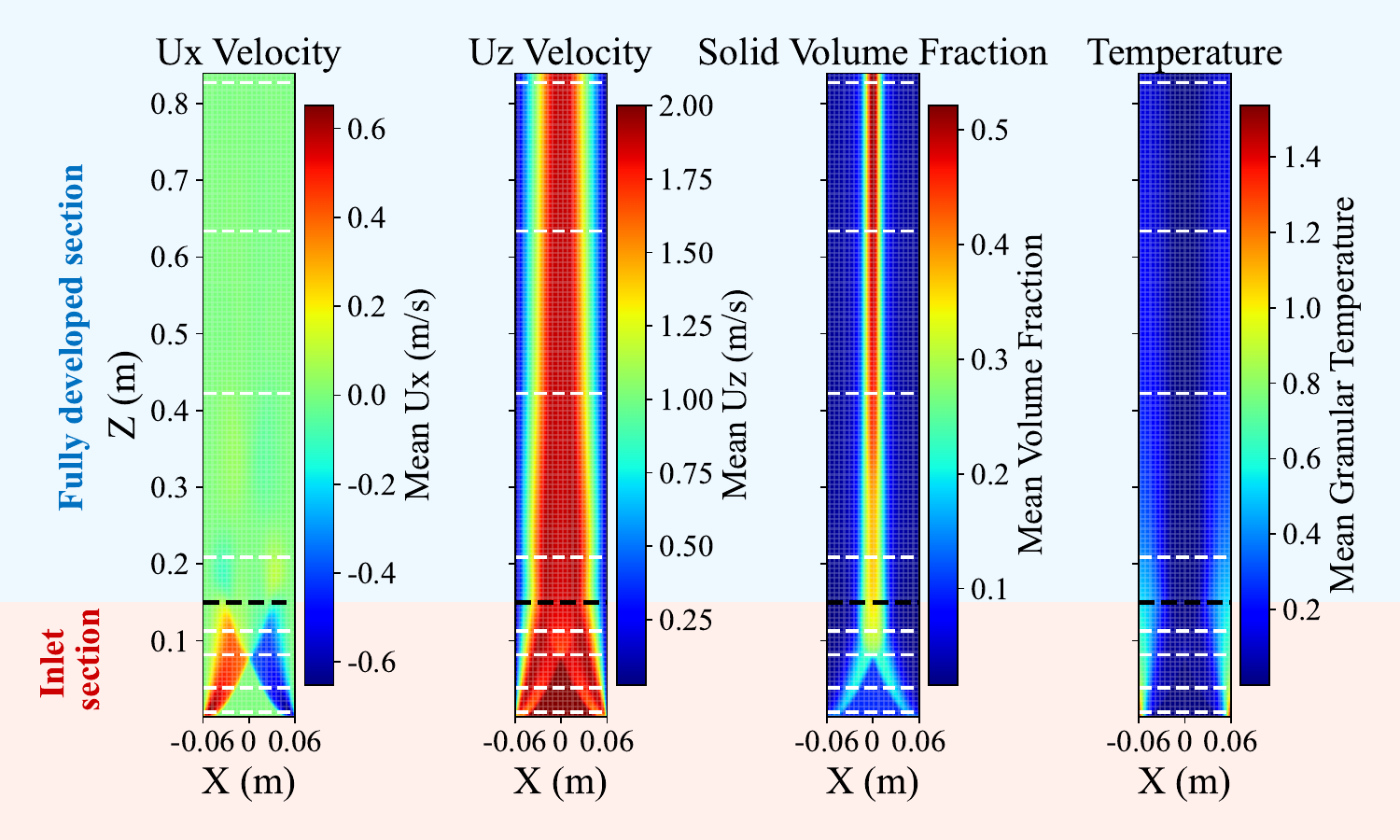}}
      \caption{Fixed-height data selection strategy} \label{2_contour_8H_data}
\end{figure}

In practice, selecting data supervision points at fixed heights is consistent with common experimental measurement procedures using for example probes, because the velocity, solid volume fraction, and granular temperature can be measured at prescribed cross-sections. In this study, $z = 0.15$ m is defined as the boundary between the inlet-development section and the developed-flow section, as indicated by the black dashed line in Figure~\ref{2_contour_8H_data}. The simulation data show that the distributions of several physical quantities change markedly at this inlet region, particularly in the x-direction velocity contour. Above this height, particle motion is almost aligned with the main flow direction. As the particles move along the pipe axis, the gradients of the physical quantities evolve continuously. To capture the flow details from inlet to outlet and reconstruct the complete flow field, four cross-sectional heights are selected within $z\le0.15$ m. In the remaining region, four additional cross-sections with relatively uniform spacing are used as data supervision points. These sampling lines pass through the center of the CFD computational grid, and the sampling points are uniformly distributed along each sampling line. These points account for approximately 1.4\% of the total data. For the inverse problem of granular pipe flow, the eight selected height locations are $z$ = [0.00675, 0.03825, 0.08175, 0.11325, 0.20925, 0.42225, 0.63375, 0.82725] m, as indicated by the white dashed lines in Figure~\ref{2_contour_8H_data}.

\begin{figure}[htbp]
\centering

\begin{minipage}{0.48\textwidth}
\centering
\includegraphics[width=\textwidth]{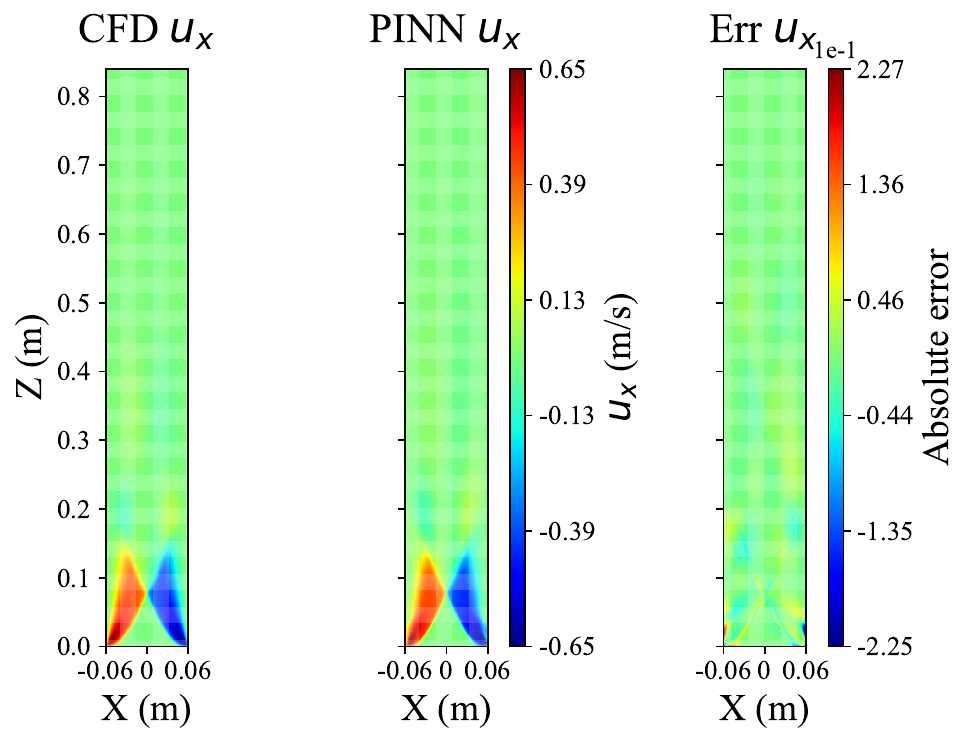}
\par(a) $u_x$
\end{minipage}
\hfill
\begin{minipage}{0.48\textwidth}
\centering
\includegraphics[width=\textwidth]{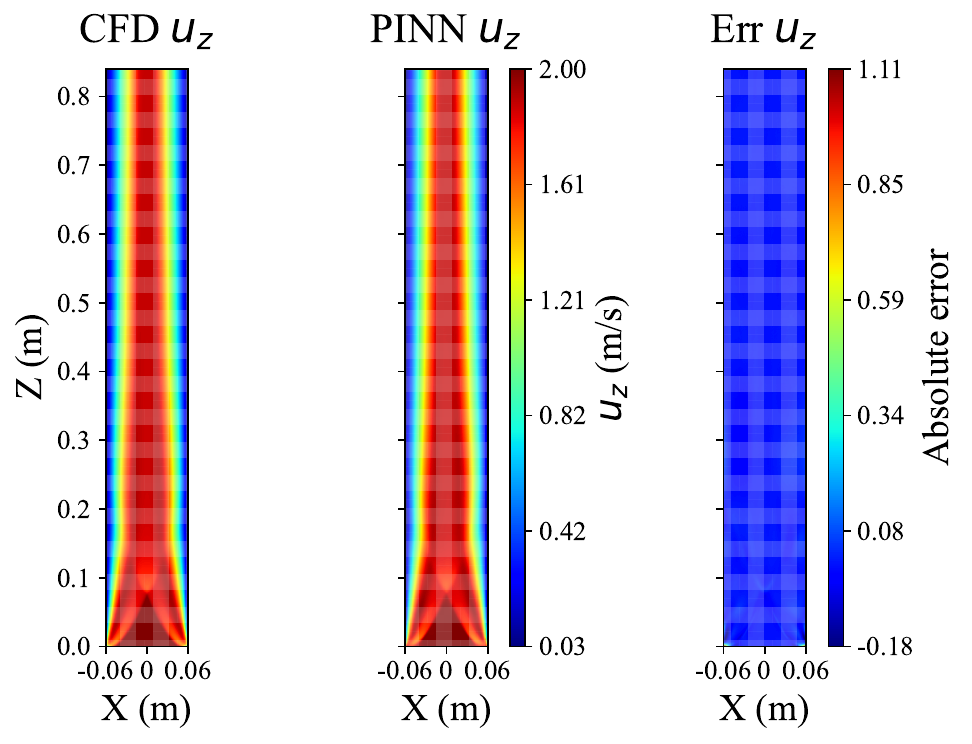}
\par(b) $u_z$
\end{minipage}

\vspace{0.3cm}

\begin{minipage}{0.48\textwidth}
\centering
\includegraphics[width=\textwidth]{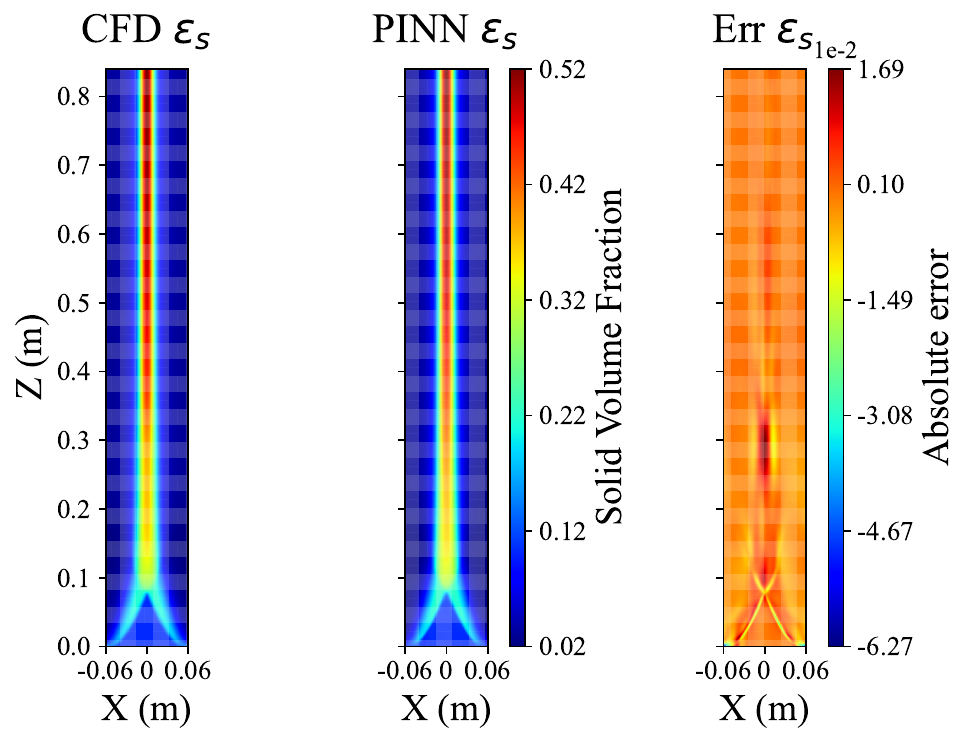}
\par(c) $\varepsilon_s$
\end{minipage}
\hfill
\begin{minipage}{0.48\textwidth}
\centering
\includegraphics[width=\textwidth]{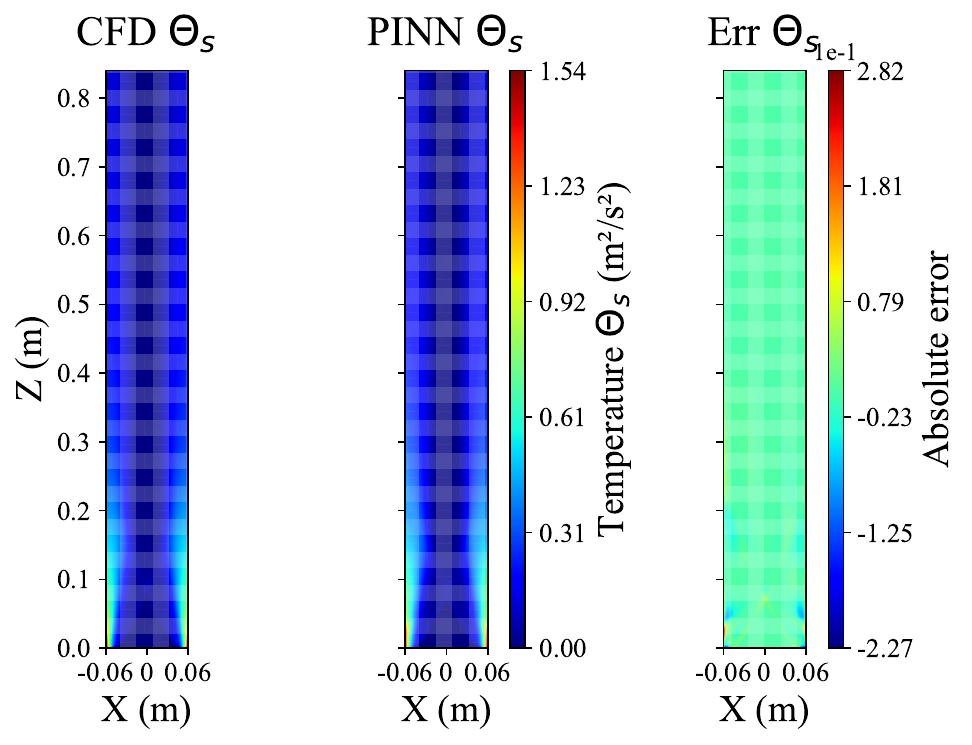}
\par(d) $\Theta_s$
\end{minipage}

\caption{Comparison between PINN-reconstructed and CFD reference flow fields under fixed-height data supervision}
\label{pinn_cfd_comparison}
\end{figure}

The flow field reconstructed by PINN from sparse data exhibits good overall agreement with the CFD simulation results, as illustrated in Figure~\ref{pinn_cfd_comparison}. Figure~\ref{pinn_cfd_comparison}(a) presents the radial velocity field, $u_x$, while Figure~\ref{pinn_cfd_comparison}(b) shows the axial velocity field, $u_z$. Both fields exhibit symmetric structures about the central axis of the pipe. Owing to the underdeveloped flow at the inlet, particle--particle and particle--wall collisions are relatively intense. Under the influence of these collisions, the particles gradually migrate toward the pipe center. As the flow develops, the radial velocity progressively decreases to a low magnitude. In the fully developed region, the particles predominantly move along the main flow direction, lateral migration diminishes, and the flow becomes stable. A high-velocity core forms at the pipe center, whereas the axial velocity decreases near the wall. The PINN reconstruction errors are primarily concentrated at the inlet, near the inlet wall, and in regions with large velocity gradients during particle aggregation. Nevertheless, the PINN effectively learns the spatially varying velocity fields. Figure~\ref{pinn_cfd_comparison}(c) presents the solid volume fraction distribution, which also exhibits a symmetric structure. Consistent with the velocity fields, the particles gradually accumulate toward the pipe center from the inlet to the outlet because of collisional effects, resulting in a gradual increase in the solid volume fraction to a maximum of $0.521$ within the computational domain. Near the wall, the particle concentration is lower, with the solid volume fraction decreasing to approximately $0.015$. The overall accuracy of the solid volume fraction field reconstructed by the PINN is high, with a maximum absolute error of only $1.69 \times 10^{-2}$ in the fully developed section. The errors are concentrated near the inlet and the wall, where the solid volume fraction gradient is steep. Figure~\ref{pinn_cfd_comparison}(d) illustrates the granular temperature distribution, which represents the intensity of random particle velocity fluctuations. The CFD results indicate that the stronger velocity gradient near the wall and the more frequent particle collisions increase the particle fluctuation energy. Consequently, the granular temperature is more pronounced in the inlet section and decreases in the central region of the pipe. The PINN reconstruction errors for the granular temperature field are mainly concentrated in the inlet section and near the wall. Upon entering the fully developed section, the flow stabilizes, the intensity of particle fluctuations decreases, and the granular temperature correspondingly declines. Overall, under this operating condition, the PINN effectively reconstructs the principal flow structures observed in the CFD results, with errors primarily originating from the inlet-development section, the near-wall boundary layer, and local regions with strong gradients.

In engineering applications of granular pipe flow, the characteristics of the fully developed flow field, which gradually forms after particles enter the pipe, are of greater interest. Compared with the inlet-development section, the physical quantities in the fully developed section become stable, and momentum exchange between particles is more thorough. The distributions of velocity, solid volume fraction, and granular temperature accurately reflect the primary flow behavior during particle transport. Transport coefficients, such as solids viscosity, solids thermal conductivity, and collisional dissipation, derived from KTGF, can characterize the dynamic evolution of the particle phase. In the subsequent analysis, axial locations within the fully developed region where no data supervision is applied are selected to compare and evaluate the PINN reconstructions against the CFD simulation results. Given that the PINN training data are limited to sparse sampling at lower heights, axial heights of $z = 0.37425$, $0.48825$, $0.68475$, and $0.74925~\mathrm{m}$ are selected for detailed analysis to further assess the model accuracy at non-data-supervised heights. These heights are located between the supervised training heights and downstream of the region where particles begin to migrate radially and converge toward the center, thereby effectively mitigating the strong influence of the inlet boundary conditions on the local flow field. This enables a more objective evaluation of the PINN model's ability to reconstruct the fully developed characteristics of granular pipe flow. By analyzing the radial profiles of axial velocity, solid volume fraction, granular temperature, and the transport coefficients calculated using KTGF, including solids viscosity, solids thermal conductivity, collisional dissipation, and solid-phase pressure, at these selected heights, the reconstruction accuracy of the PINN is further examined to validate the effectiveness of the proposed framework for reconstructing the complete granular flow field from sparse data.

\begin{figure}[htbp]
\centering

\begin{minipage}{0.48\textwidth}
\centering
\includegraphics[width=\textwidth]{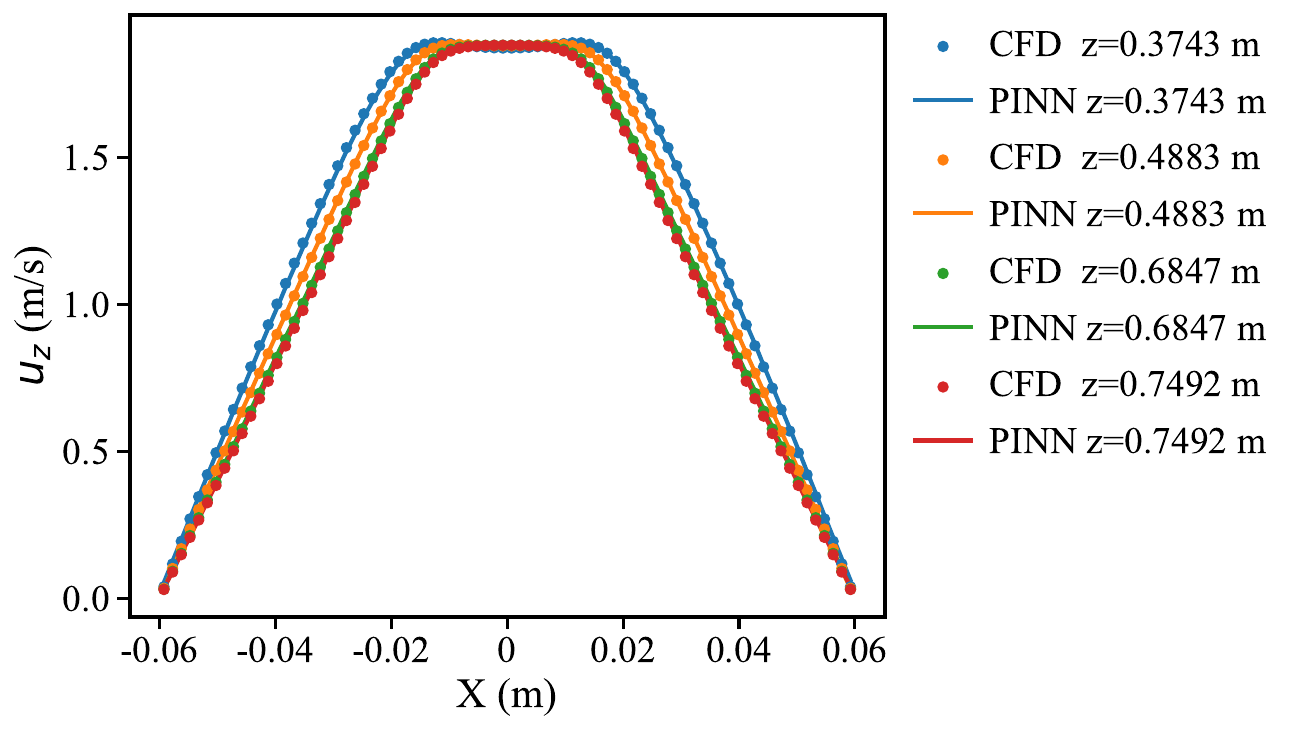}
\par(a) $u_z$
\end{minipage}
\hfill
\begin{minipage}{0.48\textwidth}
\centering
\includegraphics[width=\textwidth]{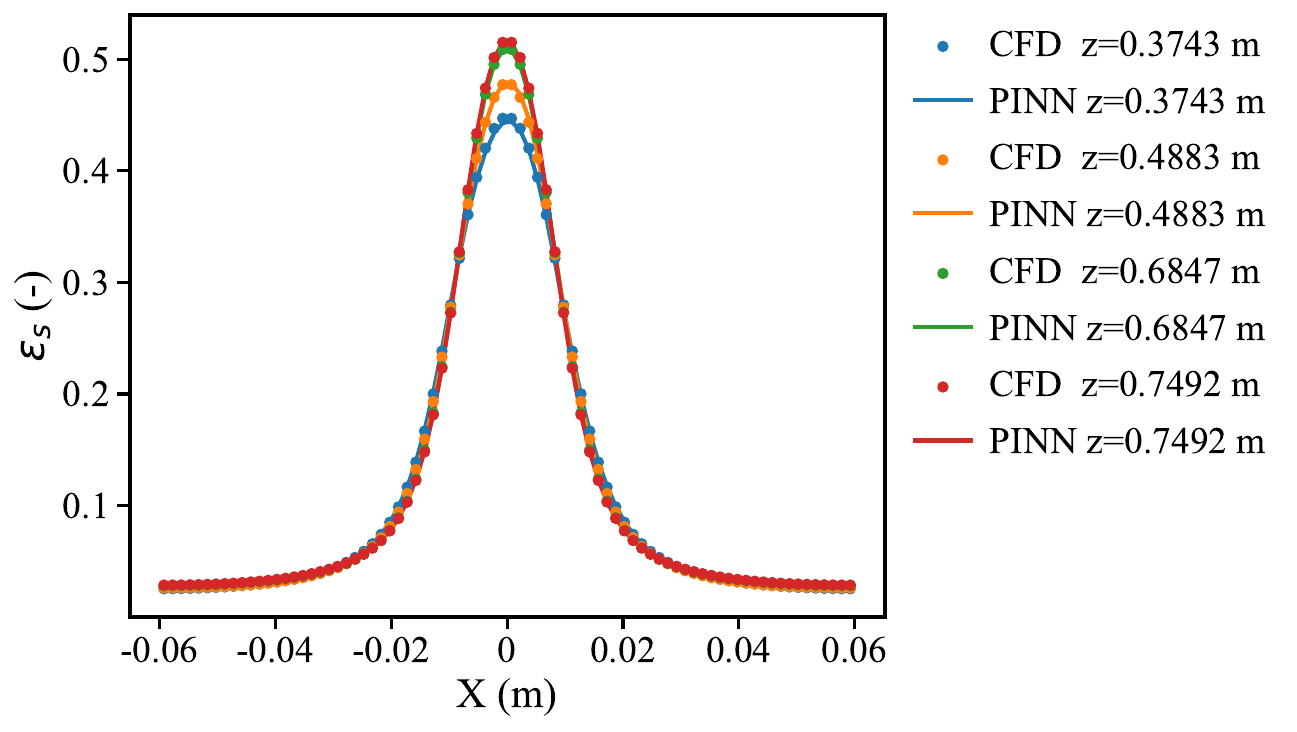}
\par(b) $\varepsilon_s$
\end{minipage}

\vspace{0.3cm}

\begin{minipage}{0.48\textwidth}
\centering
\includegraphics[width=\textwidth]{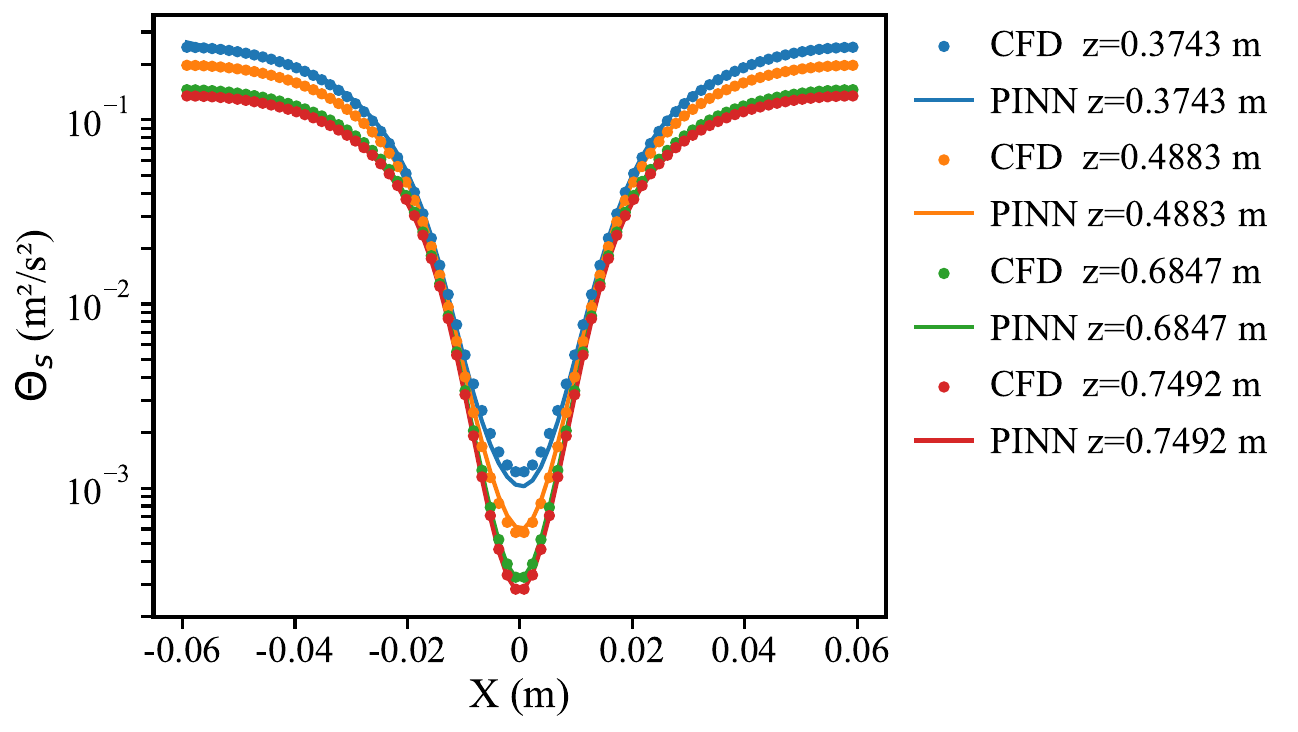}
\par(c) $\Theta_s$
\end{minipage}
\hfill
\begin{minipage}{0.48\textwidth}
\centering
\includegraphics[width=\textwidth]{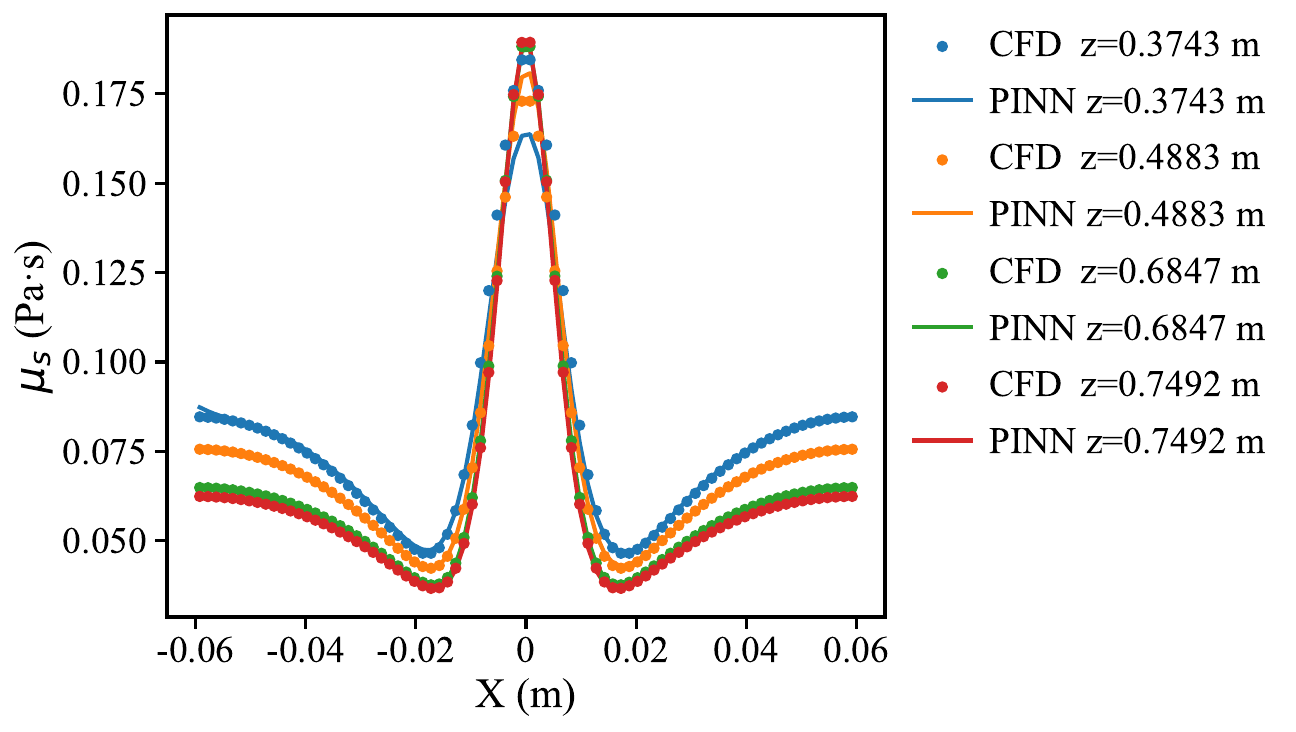}
\par(d) $\mu_s$
\end{minipage}

\vspace{0.3cm}

\begin{minipage}{0.48\textwidth}
\centering
\includegraphics[width=\textwidth]{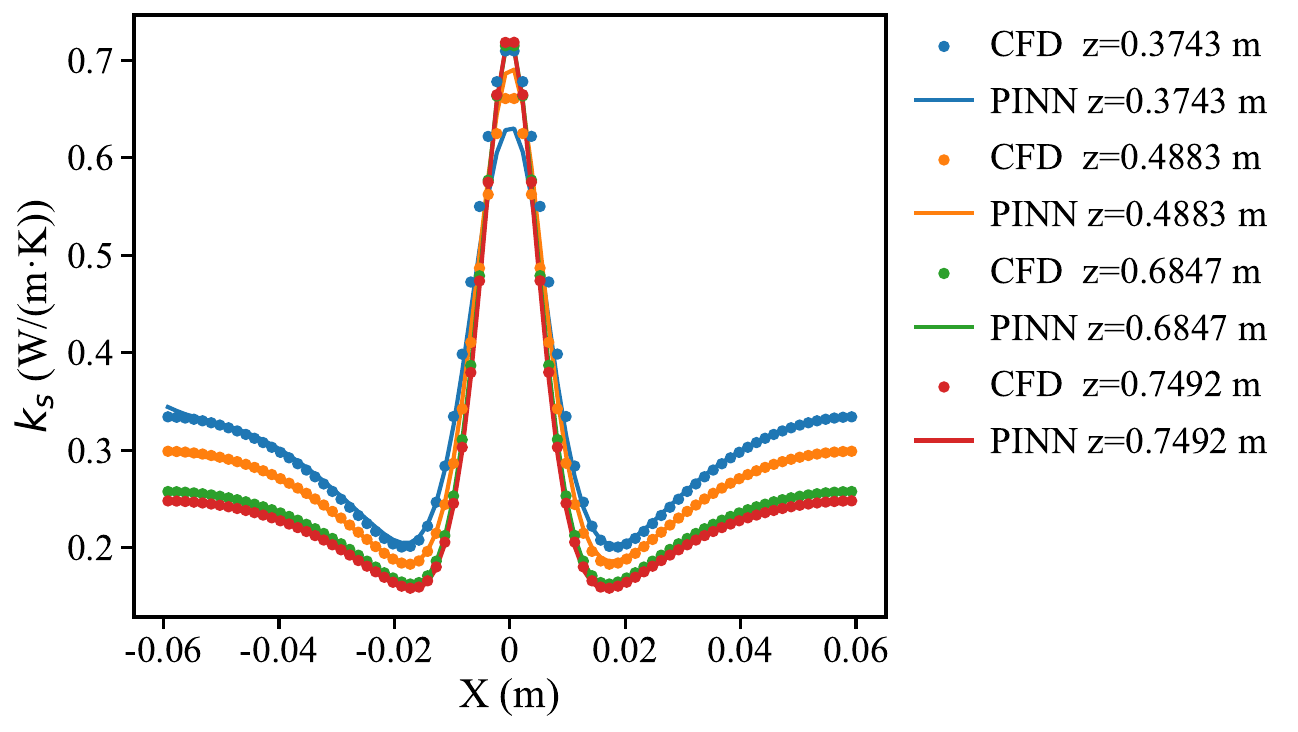}
\par(e) $k_s$
\end{minipage}
\hfill
\begin{minipage}{0.48\textwidth}
\centering
\includegraphics[width=\textwidth]{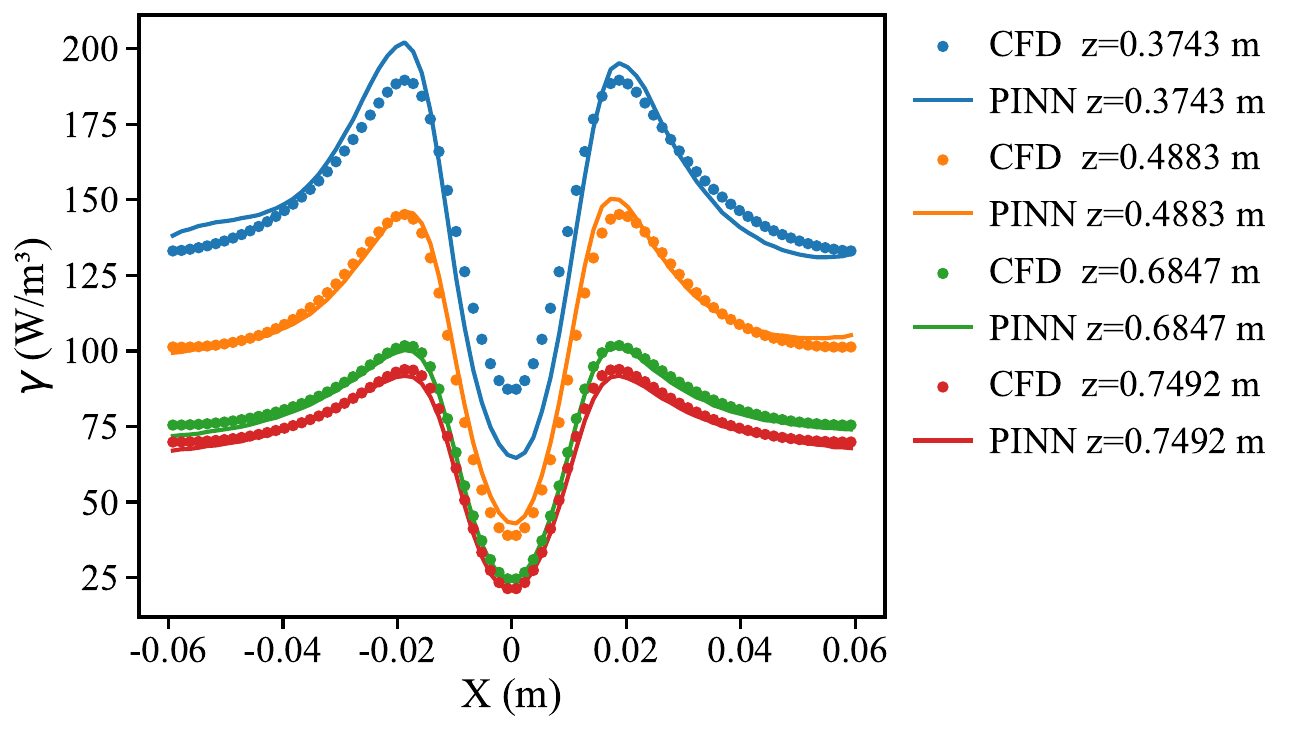}
\par(f) $\gamma$
\end{minipage}

\vspace{0.3cm}

\begin{minipage}{0.48\textwidth}
\centering
\includegraphics[width=\textwidth]{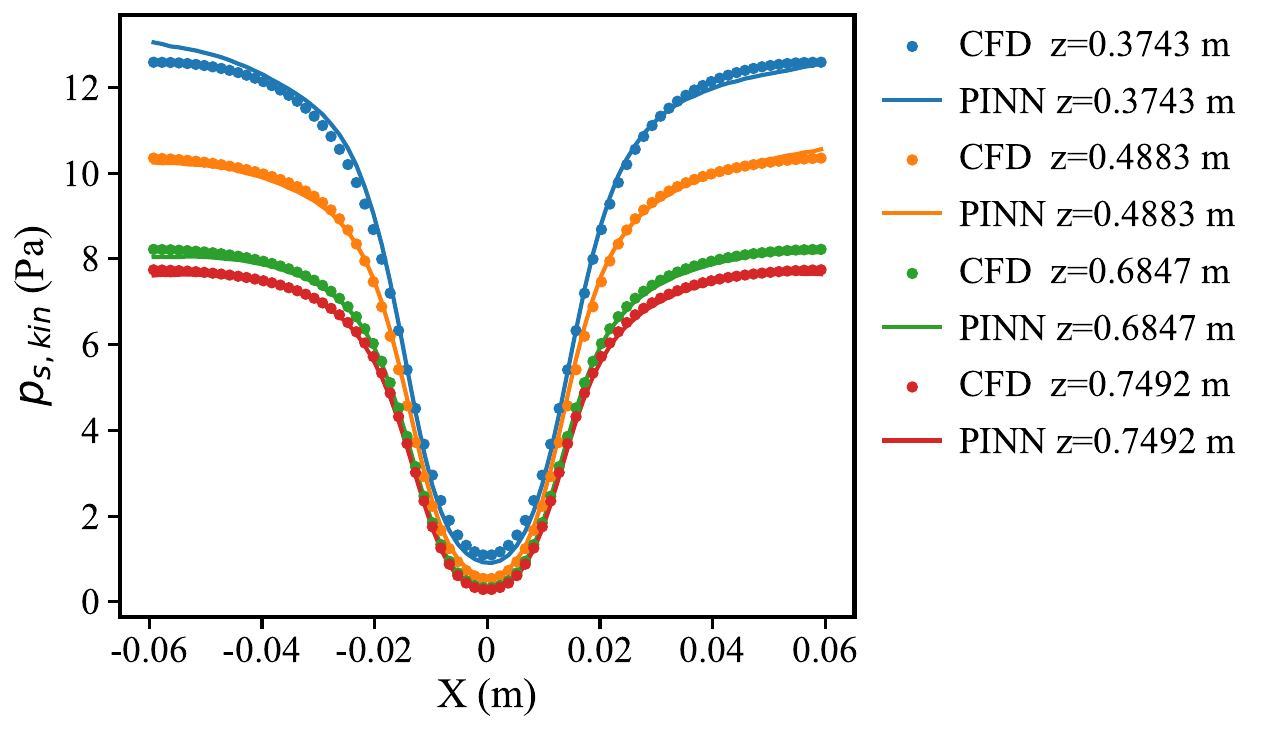}
\par(g) $p_{s,kin}$
\end{minipage}
\hfill
\begin{minipage}{0.48\textwidth}
\centering
\includegraphics[width=\textwidth]{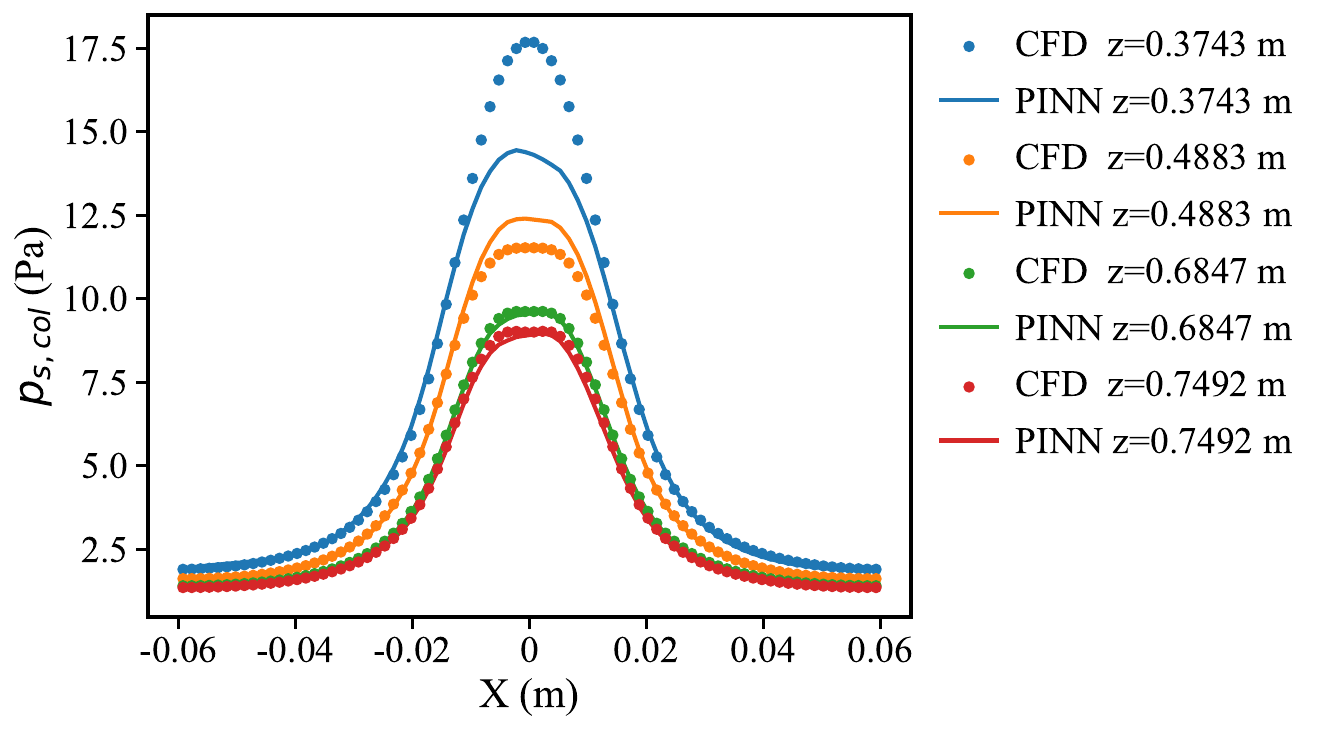}
\par(h) $p_{s,col}$
\end{minipage}

\caption{PINN predictions at selected axial locations in the fully developed region ($z=0.37425$, $0.48825$, $0.68475$, and $0.74925~\mathrm{m}$), including axial velocity $u_z$, solid volume fraction $\varepsilon_s$, and granular temperature $\Theta_s$. The reconstructed flow field was further used with KTGF constitutive relations to compute the derived quantities $\mu_s$, $k_s$, $\gamma$, $p_{s,\mathrm{kin}}$, and $p_{s,\mathrm{col}}$, which were compared with the corresponding CFD reference values.}
\label{PINN_predictions}
\end{figure}

As depicted in Figure~\ref{PINN_predictions}, the PINN predictions at the selected unsupervised analysis heights show good agreement with the CFD simulation results for the different physical quantities. Figures~\ref{PINN_predictions}(a), \ref{PINN_predictions}(b), and \ref{PINN_predictions}(c) present the axial velocity, solid volume fraction, and granular temperature at different heights, respectively, all of which exhibit symmetric distributions about the central axis. For the axial velocity, the velocity near the wall is close to zero, primarily because of wall confinement and intensified particle--wall interactions. The velocity near the axis initially increases and then approaches a stable maximum of approximately $1.88~\mathrm{m\,s^{-1}}$. As the height increases, the radial gradient of the axial velocity from the wall to the high-velocity core gradually decreases. The PINN demonstrates high accuracy in reconstructing the axial velocity in the fully developed section. For the solid volume fraction, the value near the wall is close to zero, increases along the radial direction, and reaches a peak at the pipe center. As the height increases and the particle flow stabilizes, the peak solid volume fraction gradually increases to a maximum of $0.52$. The PINN effectively reconstructs the non-uniform spatial distribution characteristic of the stabilized flow. The granular temperature exhibits a trend opposite to that of the solid volume fraction: it is low in the central region and high near the wall, with a significant order-of-magnitude variation on the logarithmic scale. This indicates that particles near the wall are more strongly affected by shear and collisions, resulting in higher particle fluctuation energy. As the height increases, the overall granular temperature decreases, and the particle flow becomes more stable. The PINN reasonably reconstructs the granular temperature distribution, which is low in the center and high near the wall, although some error remains near the central minimum.

The solids viscosity, $\mu_s$, solids thermal conductivity, $k_s$, and collisional dissipation, $\gamma$, are closely related to the solid volume fraction and granular temperature and are calculated using Eqs.~\eqref{eq:particle_viscosity},~\eqref{eq:solids_thermal_conductivity}, and~\eqref{eq:collisional_dissipation}, respectively. Specifically, Figures~\ref{PINN_predictions}(d) and~\ref{PINN_predictions}(e) present $\mu_s$ and $k_s$, respectively. Because the formulations of these two coefficients have similar nonlinear dependencies on the granular temperature and solid volume fraction, their spatial trends are completely consistent, with both decreasing slightly from the wall toward the pipe center. The values of $\mu_s$ and $k_s$ calculated from the PINN outputs show minor deviations from the CFD reference values at the central peak. Near the wall, $k_s$ and $\mu_s$ gradually decrease with increasing height, whereas their magnitudes at the center increase with height. Figure~\ref{PINN_predictions}(f) presents the collisional dissipation, whose variation trend is opposite to those of $k_s$ and $\mu_s$. From the wall toward the center, the collisional dissipation initially increases slightly and then gradually decreases. As the height increases, the flow stabilizes, the intensities of particle collisions and fluctuations decrease, and the collisional dissipation gradually diminishes. The collisional dissipation predicted by the PINN also exhibits some deviation at the central peak, with the prediction accuracy improving at greater heights. Regarding the solid-phase pressure, the kinetic pressure, $p_{s,\mathrm{kin}}$, and collisional pressure, $p_{s,\mathrm{col}}$, also exhibit symmetric distributions but opposite variation patterns. The kinetic pressure, which is more strongly affected by the granular temperature and fluctuation intensity, decreases from the wall toward the center and with increasing height. Conversely, the collisional pressure increases from the wall toward the center and decreases with increasing height. After the particles converge in the central region, the collisional pressure becomes the dominant contribution to the solid-phase pressure. The accuracy of the PINN-predicted kinetic pressure is notably higher than that of the collisional pressure. Specifically, deviations occur at the peak of the collisional pressure, but the overall error decreases with increasing height, indicating that the flow gradually stabilizes during axial development. Overall, the transport coefficients and solid-phase pressures calculated from the PINN outputs show good agreement with the CFD reference results in terms of their overall distribution trends. The PINN accurately reconstructs the physical fields and correctly represents the nonlinear coupling between the solid volume fraction and granular temperature.

\begin{figure}[htbp]
\centering
\includegraphics[width=0.48\textwidth]{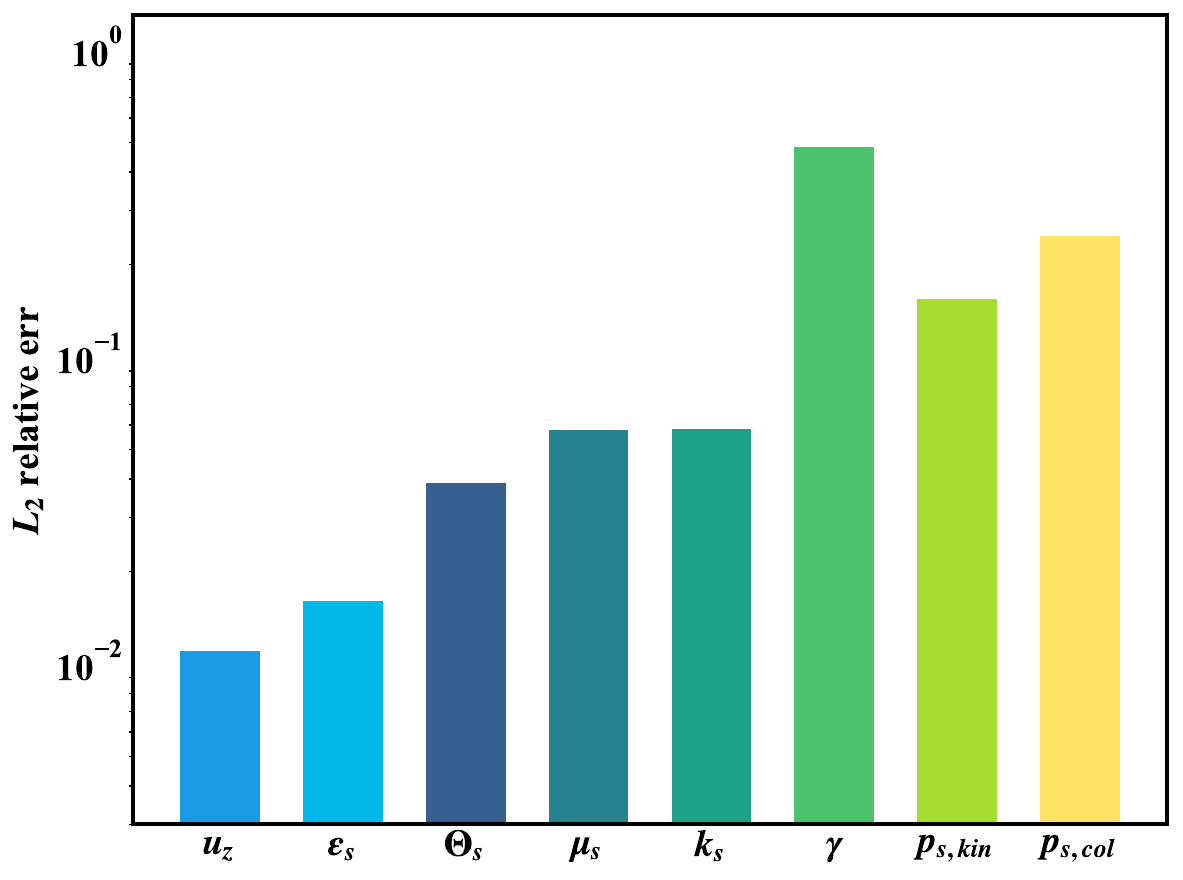}
\caption{Relative errors of all physical quantities over the full flow field}
\label{5_L2_rel_all}
\end{figure}

\begin{figure}[htbp]
\centering
\includegraphics[width=0.75\textwidth]{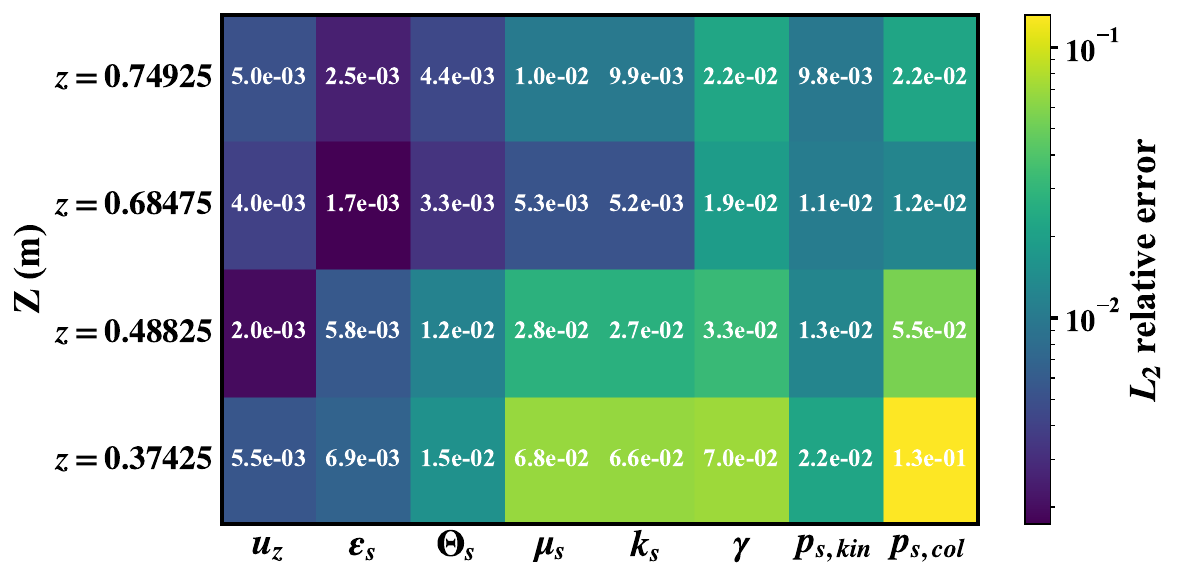}
\caption{Relative errors of all physical quantities at $z=0.37425$, $0.48825$, $0.68475$, and $0.74925~\mathrm{m}$}
\label{6_L2_rel_4z}
\end{figure}

From the perspective of error distribution, the errors for each physical quantity are primarily concentrated at the edges of regions with steep gradients, in the inlet section, and near the inlet walls, where particle collisions and shear effects are relatively strong. Despite some deviations in the PINN reconstructions in these regions, the overall trends, flow-field structures, and magnitudes of the physical quantities are well captured. As shown in Figure~\ref{5_L2_rel_all}, the full-field relative errors of the axial velocity, solid volume fraction, and granular temperature are $1.1 \times 10^{-2}$, $1.6 \times 10^{-2}$, and $3.7 \times 10^{-2}$, respectively.

For the transport coefficients derived from the solid volume fraction and granular temperature, Figure~\ref{6_L2_rel_4z} demonstrates that the relative errors decrease as the analysis height increases and the flow stabilizes. Specifically, the relative error of the axial velocity decreases from $6.9 \times 10^{-3}$ to $4.5 \times 10^{-3}$, that of the solid volume fraction decreases from $6.5 \times 10^{-3}$ to $5.7 \times 10^{-3}$, and that of the granular temperature decreases from $1.2 \times 10^{-2}$ to $4.8 \times 10^{-3}$. For the KTGF-derived transport coefficients, the prediction errors also decrease with increasing height. The error levels of the collisional dissipation, $\gamma$, and collisional pressure, $p_{s,\mathrm{col}}$, are slightly higher than those of the other quantities. In addition, the error-variation patterns of the different transport coefficients exhibit strong consistency. For example, $k_s$ and $\mu_s$ exhibit nearly identical spatial distributions and error evolution because of their similar formulations. In contrast, the collisional pressure, which is more strongly affected by particle aggregation and intense collision effects, exhibits relatively larger errors. Overall, the PINN demonstrates good physical consistency in reconstructing the complex multiphysics fields of particle-laden flow, confirming the feasibility of physics-constrained neural networks for multifield reconstruction under sparse-data conditions.

In summary, the PINN model accurately reconstructs the radial distributions of key flow-field variables, including axial velocity, solid volume fraction, and granular temperature, in granular pipe flow. Furthermore, it effectively captures the nonlinear coupling relationships among particle concentration, granular temperature, and velocity gradients, enabling the accurate derivation of transport coefficients such as solids thermal conductivity, turbulent viscosity, collisional dissipation, and solid-phase pressure. Results at different axial heights indicate that the PINN effectively learns the nonlinear interrelationships among multiple physical quantities at sections without directly applied supervisory data, successfully capturing typical flow features such as the central high-velocity zone, particle accumulation zone, near-wall high-gradient region, and areas of intense collision. Overall, the prediction accuracy of the model in the fully developed section is significantly higher than that in the inlet-development section, suggesting that the reconstruction capability of the PINN improves as the flow field gradually stabilizes.

\section{Conclusion}
This paper addresses the inverse problem of granular pipe flow with unknown inlet, outlet and wall boundary conditions by developing a PINN-based multiphysics flow-field reconstruction framework. Through a dimensionless loss formulation, physics-informed initialization, dynamic global weighting, and a locally weighted strategy for granular-temperature data loss, the proposed PINN effectively reconstructs granular-flow fields from sparse measurement data. Accurate predictions are achieved for the axial velocity, solid volume fraction, and granular temperature. In addition, the spatial distributions of solids thermal conductivity, solids viscosity, collisional dissipation, and solid-phase pressure are obtained through constitutive correlations from the kinetic theory of granular flow.
The results prove the feasibility of using PINN to reconstruct complete granular-flow fields from sparse measurement data and demonstrate the engineering potential of PINN-based reconstruction methods using fixed cross-sectional measurement data.

\section*{CRediT authorship contribution statement}
Bing Wan: writing - original draft, visualization, validation, software, methodology, investigation, formal analysis, and data curation. Bidan Zhao: Funding acquisition, writing - review and editing, supervision, methodology, investigation, formal analysis, and conceptualization. Junwu Wang: Writing - review and editing, supervision, methodology, investigation, funding acquisition, formal analysis, and conceptualization.

\section*{Acknowledgement}
This study is financially supported by the National Natural Science Foundation of China (22578485, 22378399, 22478421), and the Science Foundation of China University of Petroleum (Beijing) (2462024YJRC012, 2462024YJRC008).

\section*{Declaration of competing interest}
The authors declare that they have no known competing financial interests or personal relationships that could have appeared to influence the work reported in this paper.

\section*{Data Availability and Reproducibility Statement}
Data will be made available on request.

\section*{ORCID}
\begin{verbatim}
Bidan Zhao  https://orcid.org/0000-0003-3390-1801
Junwu Wang  https://orcid.org/0000-0003-3988-1477
\end{verbatim}

\section*{Supplementary Materials}
\section*{S1. Random data supervision points}

During the solution of the PINN inverse problem, the method used to select data supervision points significantly affects the accuracy of the flow-field reconstruction. The reconstruction accuracy varies among different sampling strategies for the data supervision points. While the preceding discussion primarily focused on a fixed-height sampling method, this section examines the random sampling approach. Although studies of granular pipe flow often focus on the reconstruction accuracy in the fully developed region, the inlet section also requires accurate reconstruction because it is characterized by underdeveloped flow, high particle fluctuation intensity, strong shear, large spatial gradients during particle migration and aggregation, and pronounced flow-field variations. To effectively reconstruct the entire flow field and facilitate comparison with the previous results, the total number of supervision points is maintained at 640. Consistent with the distribution used for fixed-height supervision, 50\% of the supervision points are randomly distributed in the region of $z \leq 0.15~\mathrm{m}$, while the remaining 50\% are distributed throughout the rest of the computational domain, thereby enhancing the supervision of the dynamic flow evolution in the inlet region.

\begin{figure}[htbp]
\centering

\begin{minipage}{0.48\textwidth}
\centering
\includegraphics[width=\textwidth]{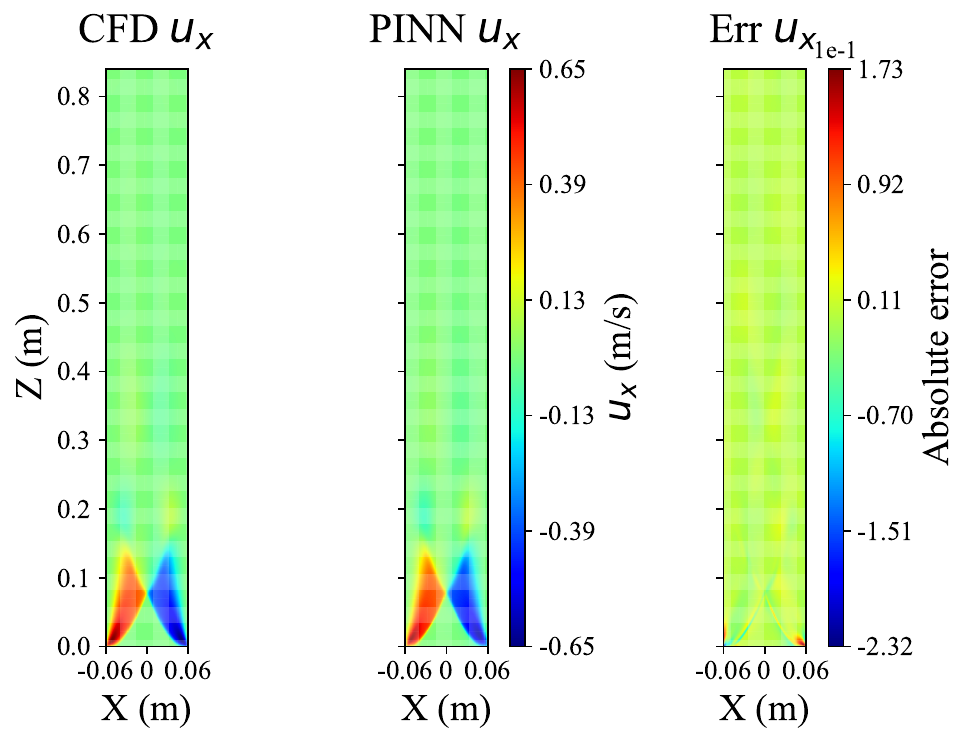}
\par(a) $u_x$
\end{minipage}
\hfill
\begin{minipage}{0.48\textwidth}
\centering
\includegraphics[width=\textwidth]{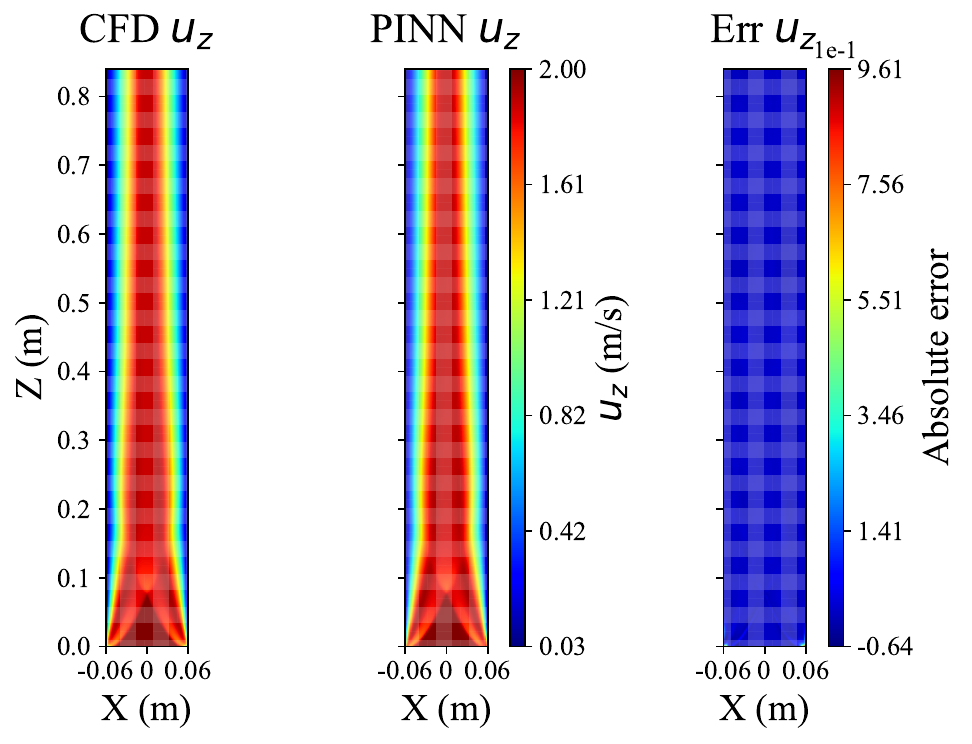}
\par(b) $u_z$
\end{minipage}

\vspace{0.3cm}

\begin{minipage}{0.48\textwidth}
\centering
\includegraphics[width=\textwidth]{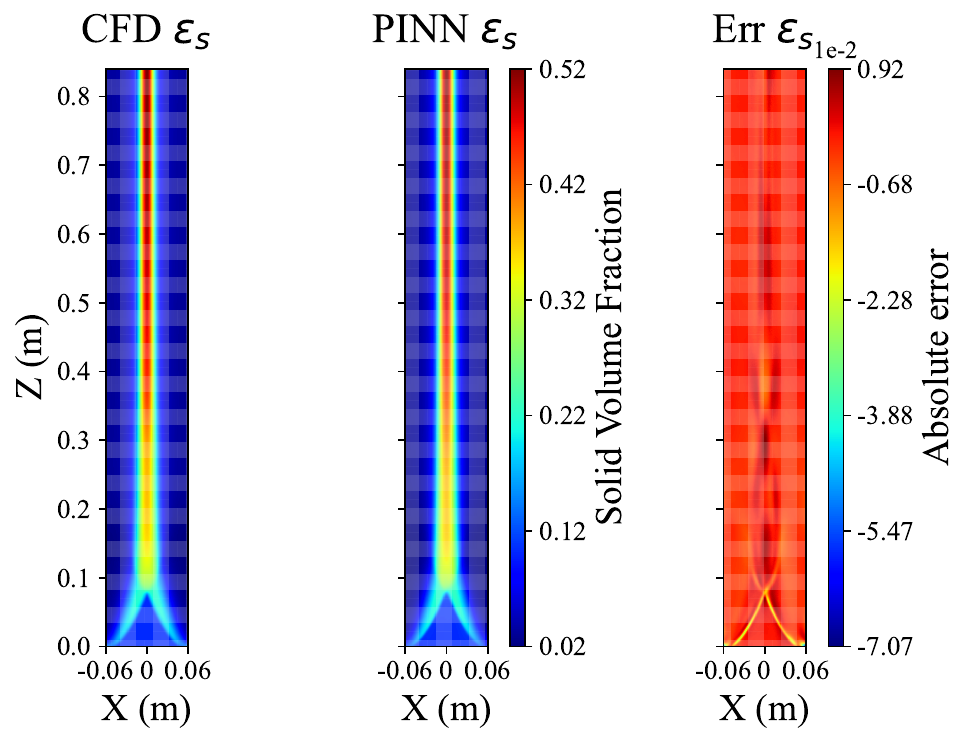}
\par(c) $\varepsilon_s$
\end{minipage}
\hfill
\begin{minipage}{0.48\textwidth}
\centering
\includegraphics[width=\textwidth]{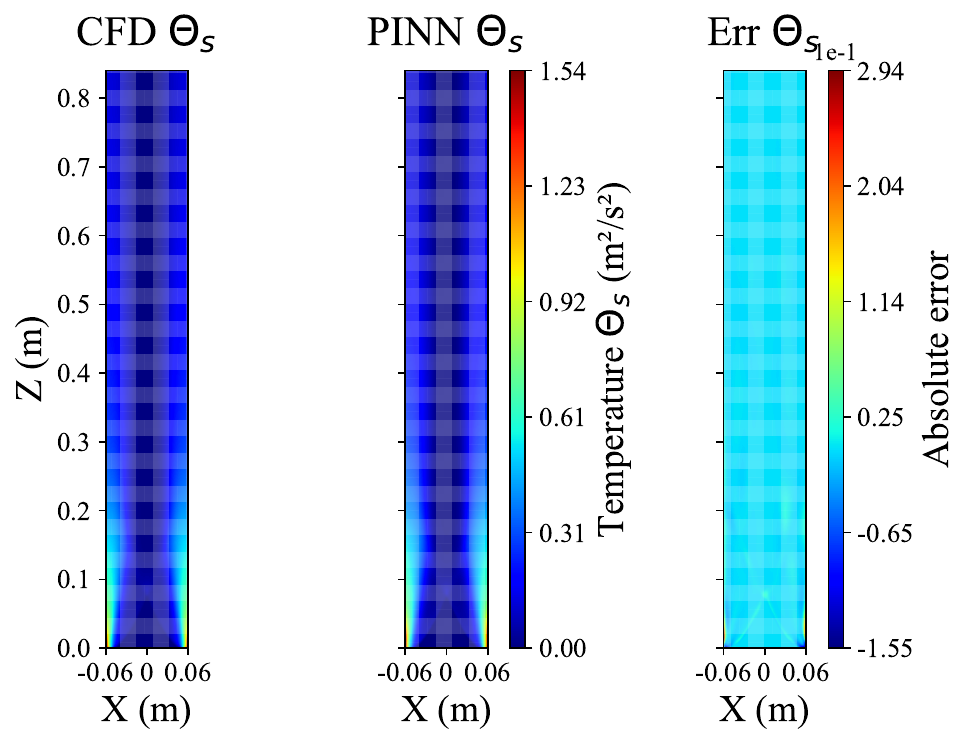}
\par(d) $\Theta_s$
\end{minipage}

\caption{Comparison between PINN-reconstructed and CFD reference flow fields under random data supervision}
\label{Comparison_between_PINN_CFD}
\end{figure}

As shown in Figure~\ref{Comparison_between_PINN_CFD}, even with randomly sampled data supervision points, the PINN effectively reconstructs the entire flow field and captures the flow patterns from the inlet to the outlet, including the non-uniform distribution induced by particle migration and aggregation. The contour and error-contour plots of the radial velocity, $u_x$, axial velocity, $u_z$, solid volume fraction, $\varepsilon_s$, and granular temperature, $\Theta_s$, exhibit a relatively uniform overall error distribution, similar to that obtained using fixed-height sampling. Regions with relatively large errors are mainly located in the inlet-development section, regions with significant spatial-gradient variations, and near the inlet wall. Figure~\ref{Comparison_between_PINN_CFD}(d) indicates that the granular temperature calculated by the PINN is higher near the right wall and lower in the outlet region.

\begin{figure}[htbp]
\centering

\begin{minipage}{0.48\textwidth}
\centering
\includegraphics[width=\textwidth]{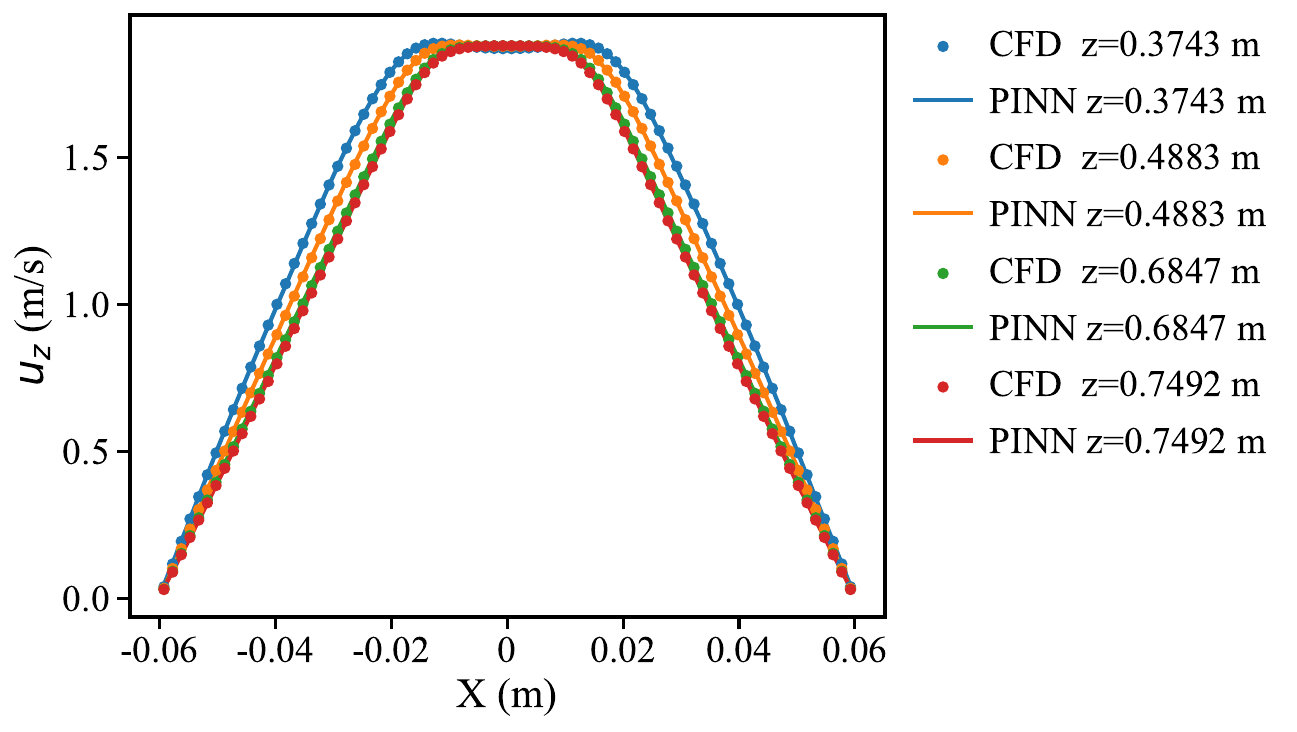}
\par(a) $u_z$
\end{minipage}
\hfill
\begin{minipage}{0.48\textwidth}
\centering
\includegraphics[width=\textwidth]{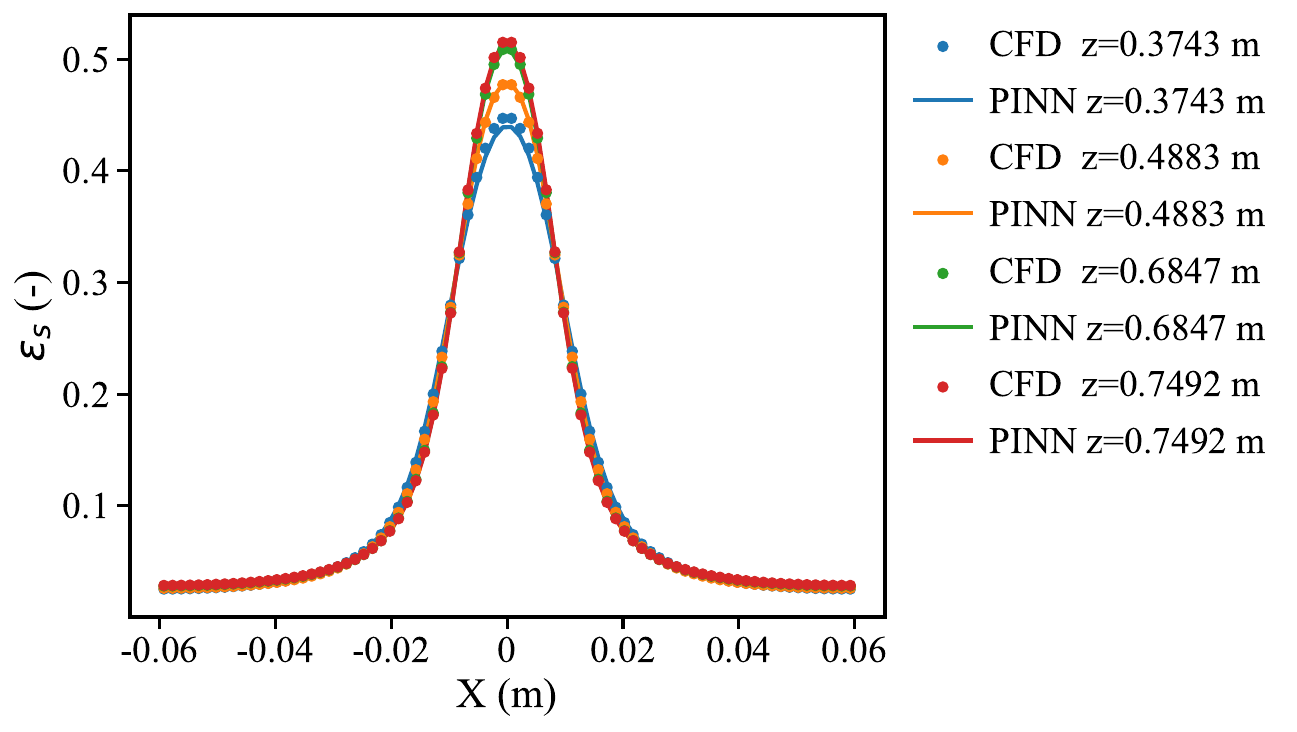}
\par(b) $\varepsilon_s$
\end{minipage}

\vspace{0.3cm}

\begin{minipage}{0.48\textwidth}
\centering
\includegraphics[width=\textwidth]{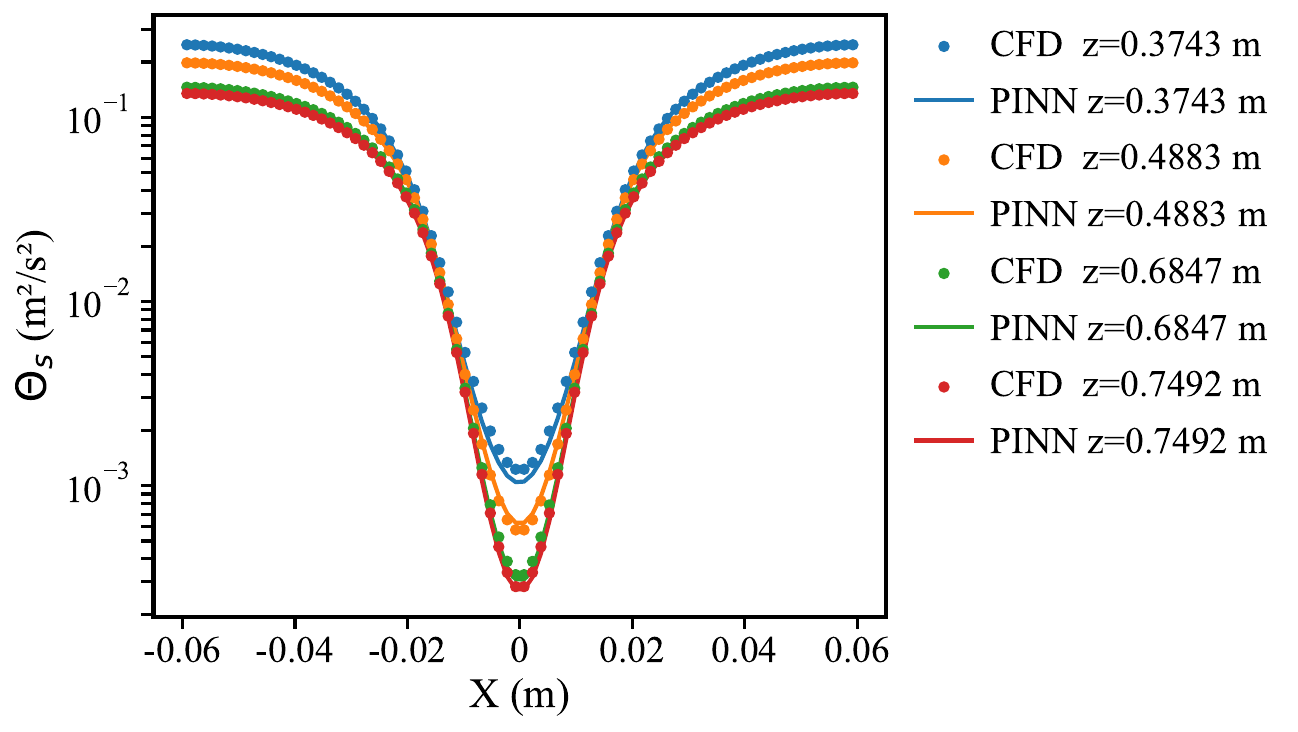}
\par(c) $\Theta_s$
\end{minipage}
\hfill
\begin{minipage}{0.48\textwidth}
\centering
\includegraphics[width=\textwidth]{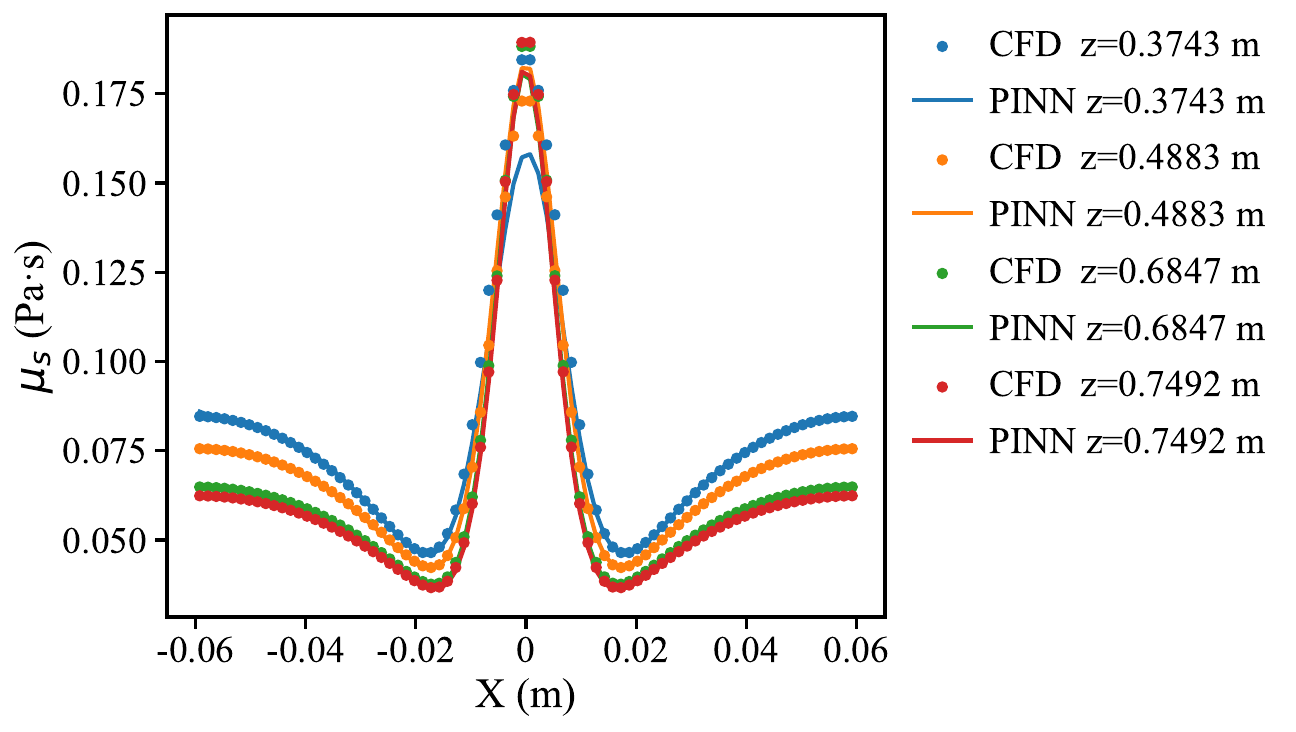}
\par(d) $\mu_s$
\end{minipage}

\vspace{0.3cm}

\begin{minipage}{0.48\textwidth}
\centering
\includegraphics[width=\textwidth]{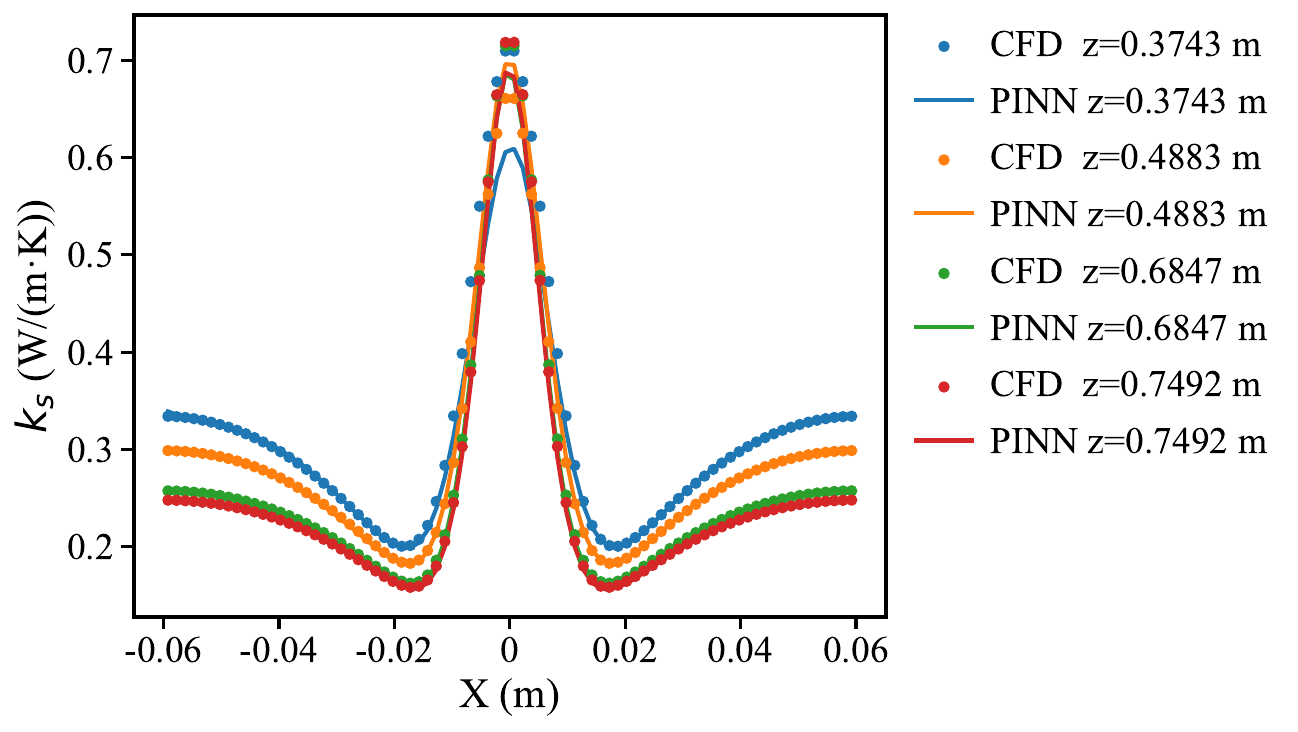}
\par(e) $k_s$
\end{minipage}
\hfill
\begin{minipage}{0.48\textwidth}
\centering
\includegraphics[width=\textwidth]{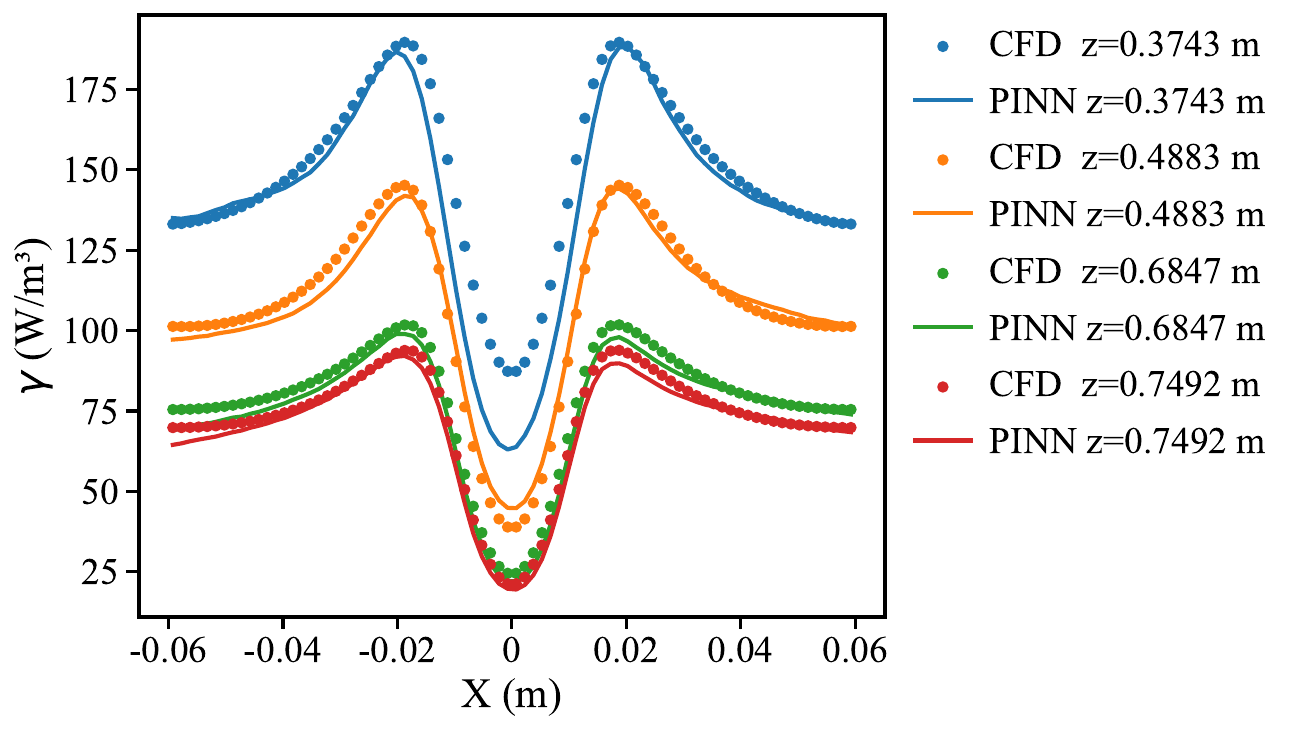}
\par(f) $\gamma$
\end{minipage}

\vspace{0.3cm}

\begin{minipage}{0.48\textwidth}
\centering
\includegraphics[width=\textwidth]{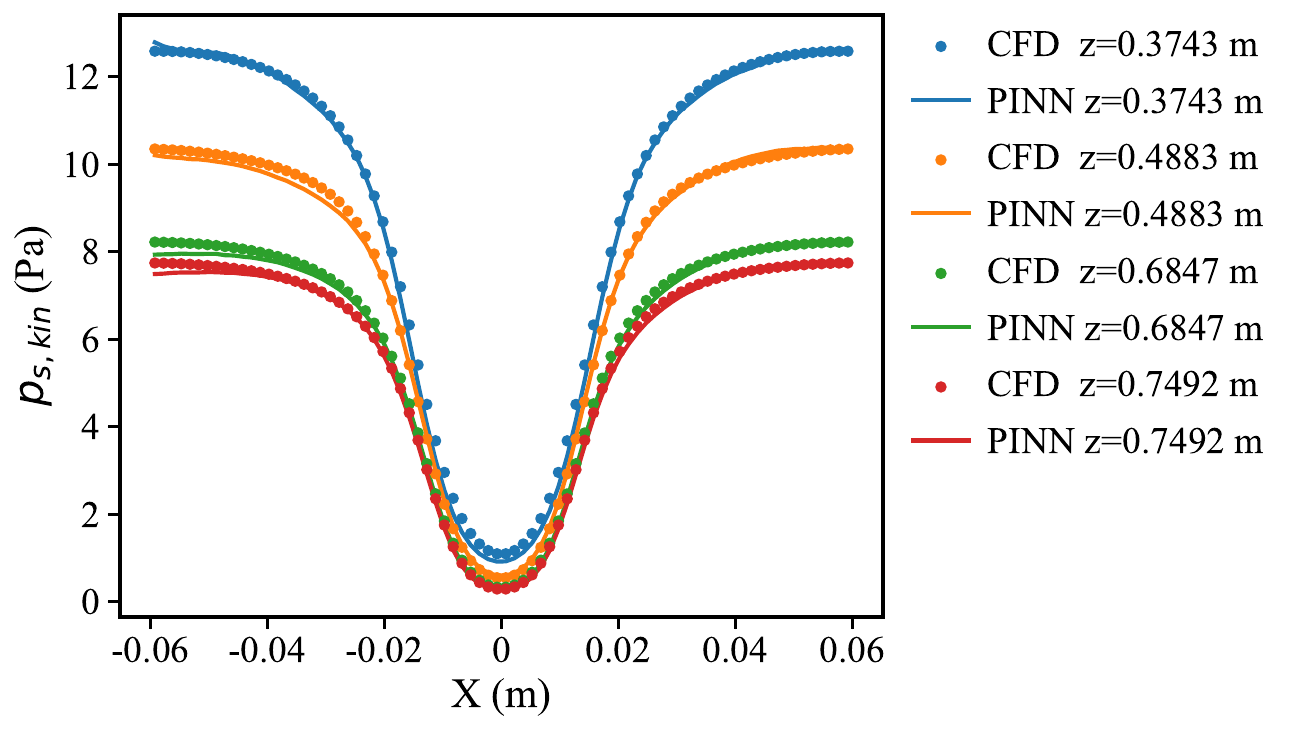}
\par(g) $p_{s,kin}$
\end{minipage}
\hfill
\begin{minipage}{0.48\textwidth}
\centering
\includegraphics[width=\textwidth]{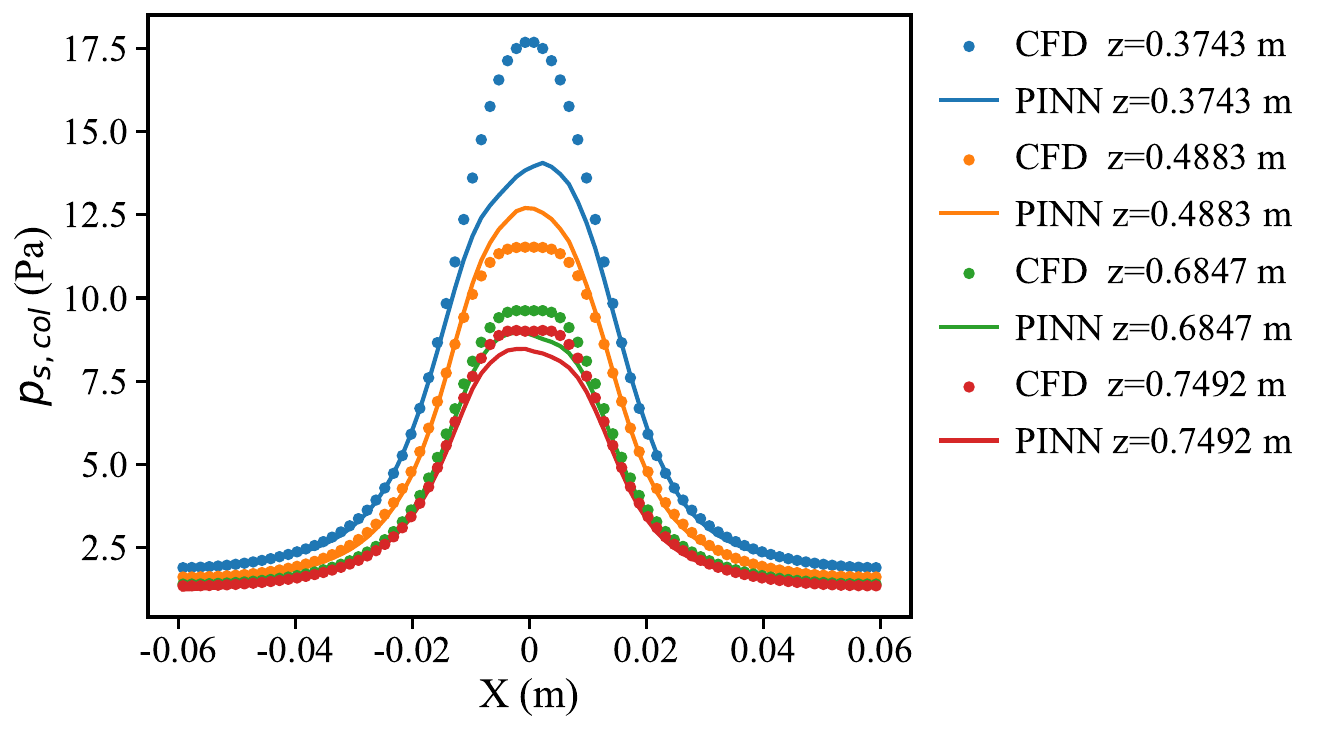}
\par(h) $p_{s,col}$
\end{minipage}

\caption{PINN predictions in the fully developed region under random sampling at $z=0.37425$, $0.48825$, $0.68475$, and $0.74925~\mathrm{m}$, including axial velocity $u_z$, solid volume fraction $\varepsilon_s$, and granular temperature $\Theta_s$. The reconstructed flow field was further used with KTGF constitutive relations to compute $\mu_s$, $k_s$, $\gamma$, $p_{s,\mathrm{kin}}$, and $p_{s,\mathrm{col}}$, which were compared with the corresponding CFD reference values.}
\label{PINN_predictions_2}
\end{figure}

The radial distributions of the various physical quantities at different heights are presented in Figure~\ref{PINN_predictions_2}. The PINN predictions show good overall agreement with the CFD results and accurately reconstruct the principal distribution characteristics at axial heights without direct data supervision. Figures~\ref{PINN_predictions_2}(a) and \ref{PINN_predictions_2}(b) indicate that the predicted velocity and solid volume fraction are highly accurate, with the predicted values nearly coinciding with the CFD reference values. As the height increases, the peak solid volume fraction gradually increases to $0.52$, the flow field stabilizes, and the particles increasingly accumulate toward the pipe center, forming a non-uniform distribution. Figure~\ref{PINN_predictions_2}(c) indicates that particles near the wall experience stronger shear and possess higher fluctuation energy. As the height increases, the overall granular temperature decreases, and the particle flow gradually stabilizes. The logarithmic temperature profiles reveal that the PINN prediction errors are mainly concentrated in the central low-temperature region, where some deviations from the CFD reference values remain.

\begin{figure}[htbp]
\centering

\begin{minipage}[t]{0.48\textwidth}
\vspace{0pt}
\centering
\includegraphics[width=\textwidth]{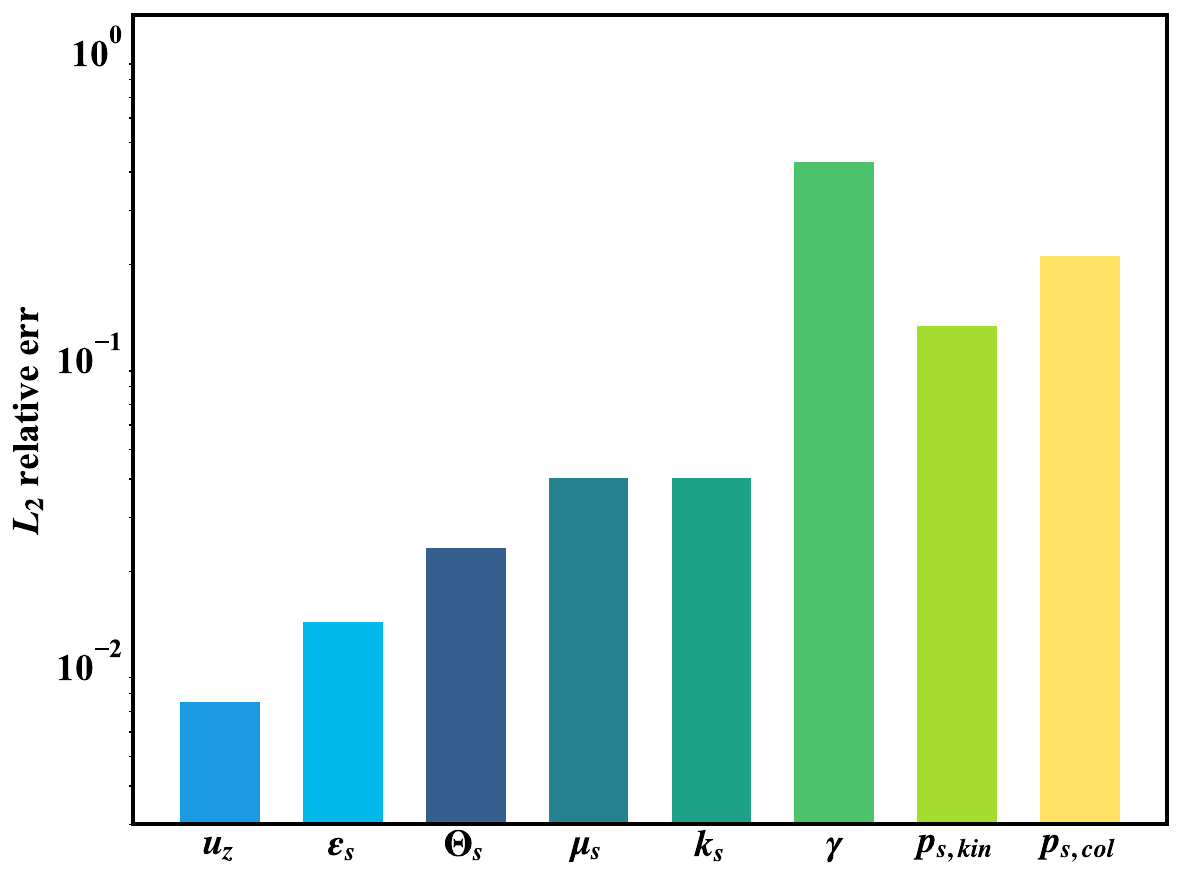}
\caption{Relative errors of physical quantities over the full flow field under random data supervision}
\label{10_L2_rel_all}
\end{minipage}
\hfill
\begin{minipage}[t]{0.48\textwidth}
\vspace{0pt}
\centering
\includegraphics[width=\textwidth]{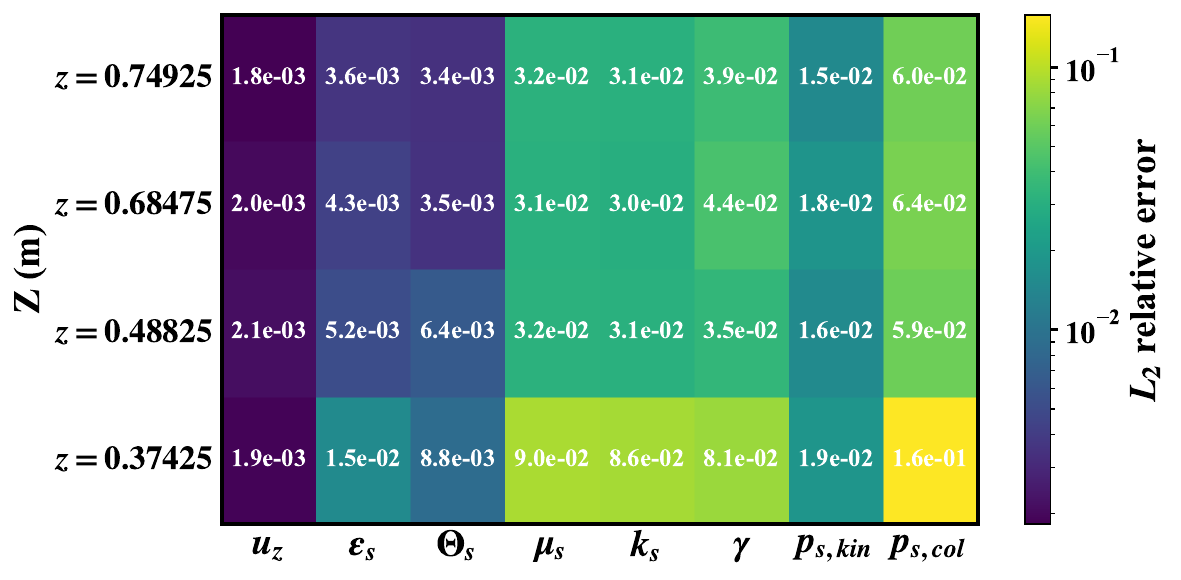}
\caption{Relative errors of physical quantities at $z=0.37425$, $0.48825$, $0.68475$, and $0.74925~\mathrm{m}$}
\label{11_L2_rel_4z}
\end{minipage}

\end{figure}

Figure~\ref{10_L2_rel_all} presents the full-field relative errors of the axial velocity, solid volume fraction, and granular temperature, which are $7.5 \times 10^{-3}$, $1.4 \times 10^{-2}$, and $2.4 \times 10^{-2}$, respectively. As shown in Figure~\ref{11_L2_rel_4z}, the relative errors of the axial velocity, solid volume fraction, and granular temperature remain relatively low and gradually decrease with increasing height. The relative error of the axial velocity remains approximately $2.0 \times 10^{-3}$, indicating high prediction accuracy. The relative error of the solid volume fraction decreases from $1.5 \times 10^{-2}$ to $3.6 \times 10^{-3}$, while that of the granular temperature decreases from $8.8 \times 10^{-3}$ to $3.4 \times 10^{-3}$, which is lower than the error obtained using the fixed-height data-supervision strategy. The KTGF-derived coefficients, including the solids thermal conductivity, $k_s$, solids viscosity, $\mu_s$, and collisional dissipation, $\gamma$, exhibit lower errors near the wall than at the pipe center, although deviations remain at the extrema of these transport coefficients. The errors of these three coefficients are comparable and decrease from approximately $8.6 \times 10^{-2}$ to $3.1 \times 10^{-2}$. Regarding the solid-phase pressure, the kinetic pressure agrees well with the CFD reference results, whereas the collisional pressure exhibits larger errors in the particle-convergence region at the pipe center. The relative error of $p_{s,\mathrm{kin}}$ decreases from $1.9 \times 10^{-2}$ to $1.5 \times 10^{-2}$, while that of $p_{s,\mathrm{col}}$ decreases from $1.2 \times 10^{-1}$ to $6.0 \times 10^{-2}$. Overall, the error of $p_{s,\mathrm{col}}$ remains higher than that of $p_{s,\mathrm{kin}}$.

Overall, under both data-supervision-point arrangement strategies, the PINN effectively reconstructs the main flow structures and multiphysics-field distributions. The prediction accuracy for the axial velocity and solid volume fraction is generally high, accurately capturing typical flow features such as the central high-velocity zone, the non-uniform particle distribution, and near-wall gradient variations during flow development in the pipe. In contrast, variables that are more sensitive to local collisions, shear, and particle migration, including granular temperature, solids viscosity, solids thermal conductivity, collisional dissipation, and solid-phase pressure, exhibit relatively larger errors. These errors are primarily concentrated in regions with intense flow variations, such as the inlet-development section, near-wall regions, and central peak regions. Both supervision strategies confirm the feasibility of the PINN for reconstructing multiphysics information in granular pipe flow from sparse data. Owing to its broader spatial coverage and stronger global constraint capability, the random sampling strategy exhibits superior stability and overall reconstruction accuracy for complex flow fields.

\section*{S2. Effect of the number of data supervision points on reconstruction accuracy}
In practical engineering and experimental measurements, flow-field information can typically be obtained only from a limited number of measurement sections. Consequently, the quantity and spatial distribution of data supervision points directly affect the reconstruction capability of the PINN. As discussed previously, both fixed-height and random sampling methods can yield accurate reconstructions. Moreover, acquiring data at fixed heights is more consistent with typical experimental configurations, in which measurement sections are often arranged at specific axial positions. Therefore, investigating the effect of varying the number of measurement heights on the PINN reconstruction accuracy is crucial for assessing the applicability of this method under sparse experimental data conditions.

\begin{figure}[htbp]
      \centerline{\includegraphics[width=0.75\textwidth]{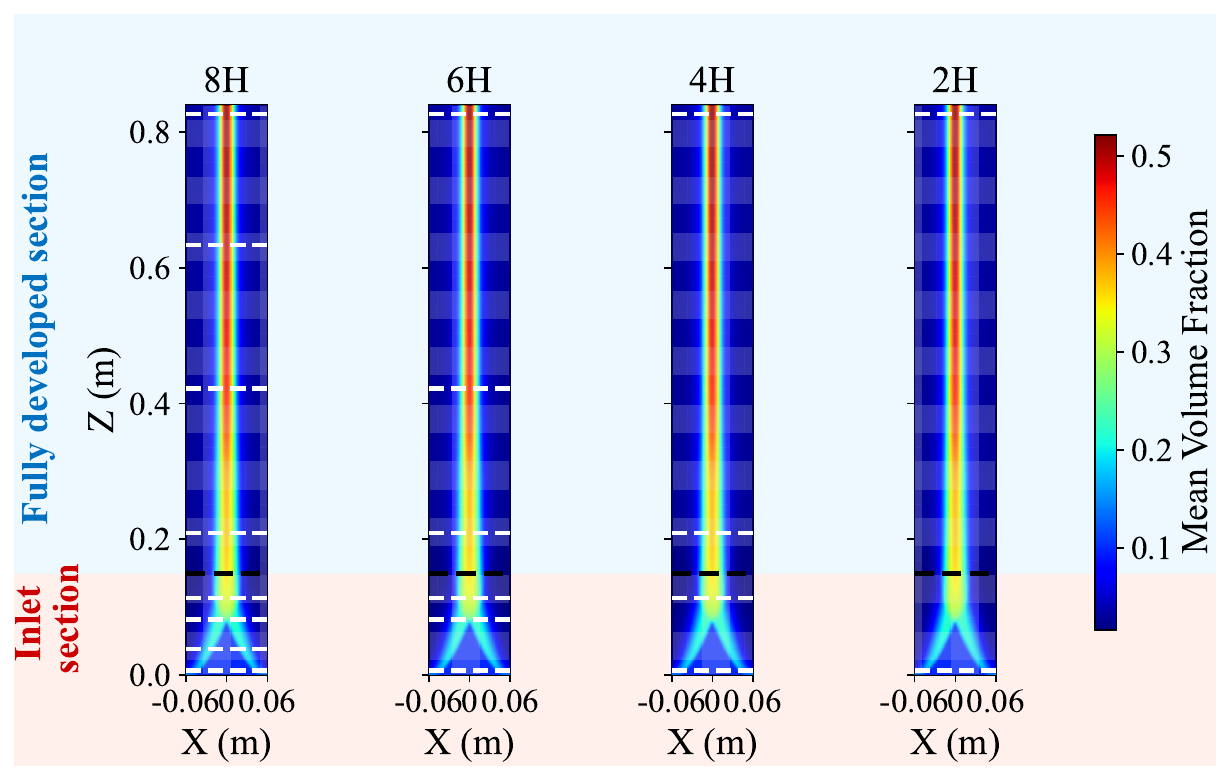}}
      \caption{Fixed-height data selection strategy} \label{13_contour_8642H_data}
\end{figure}

To further analyze the effect of reducing the amount of supervision data on the PINN solution capability, the network architecture, weight initialization, and loss function were maintained consistently with those used in the previous section, while data from 8, 6, 4, and 2 different axial heights were selected as supervision information. A progressively sparser sampling strategy was adopted by gradually reducing the number of supervised heights to reconstruct the flow field under different supervision-data volumes. As shown in Figure~\ref{13_contour_8642H_data}, the data-supervision heights in the inlet and fully developed sections were gradually reduced, ultimately retaining only the two extreme monitoring sections nearest the inlet and outlet. By progressively decreasing the section density, the reconstruction capability of the PINN for the entire flow field under diminishing supervision information was systematically evaluated. By comparing the full-field relative errors of each physical quantity under different data volumes and the radial-distribution errors at the previously selected analysis heights, the effects of sparse supervision data on reconstruction accuracy and model generalization were quantitatively assessed, thereby providing guidance for future experiments.

\begin{figure}[htbp]
\centering

\begin{minipage}{0.48\textwidth}
\centering
\includegraphics[width=\textwidth]{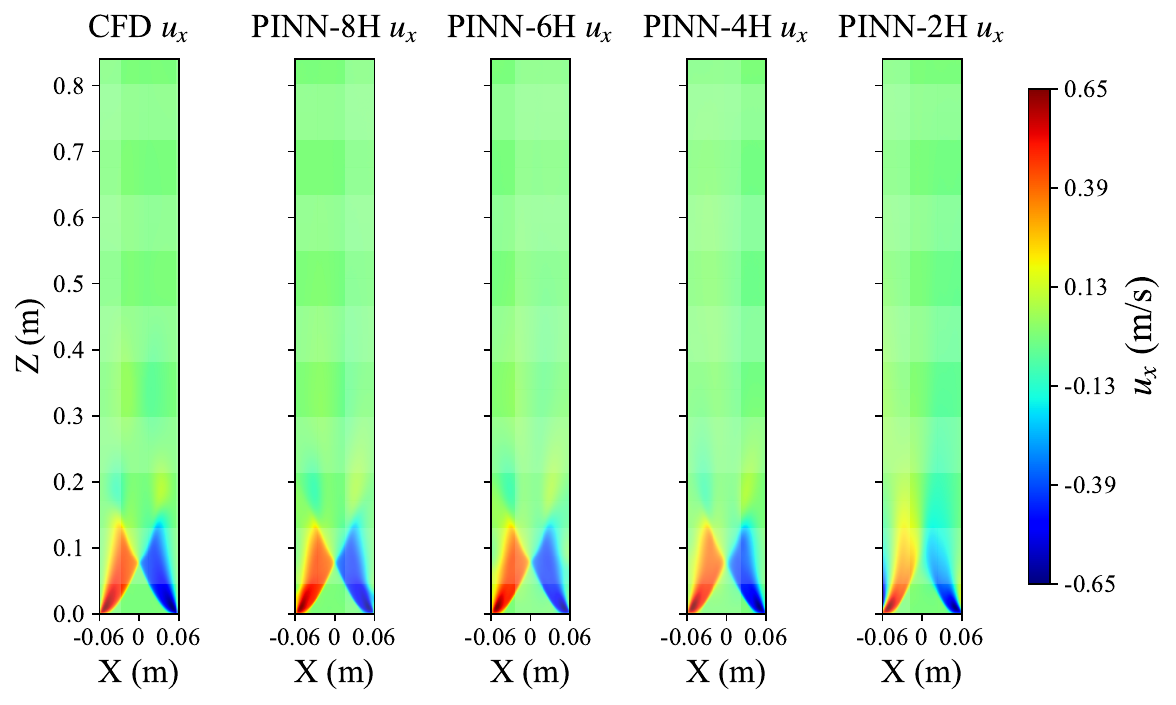}
\par(a) $u_x$
\end{minipage}
\hfill
\begin{minipage}{0.48\textwidth}
\centering
\includegraphics[width=\textwidth]{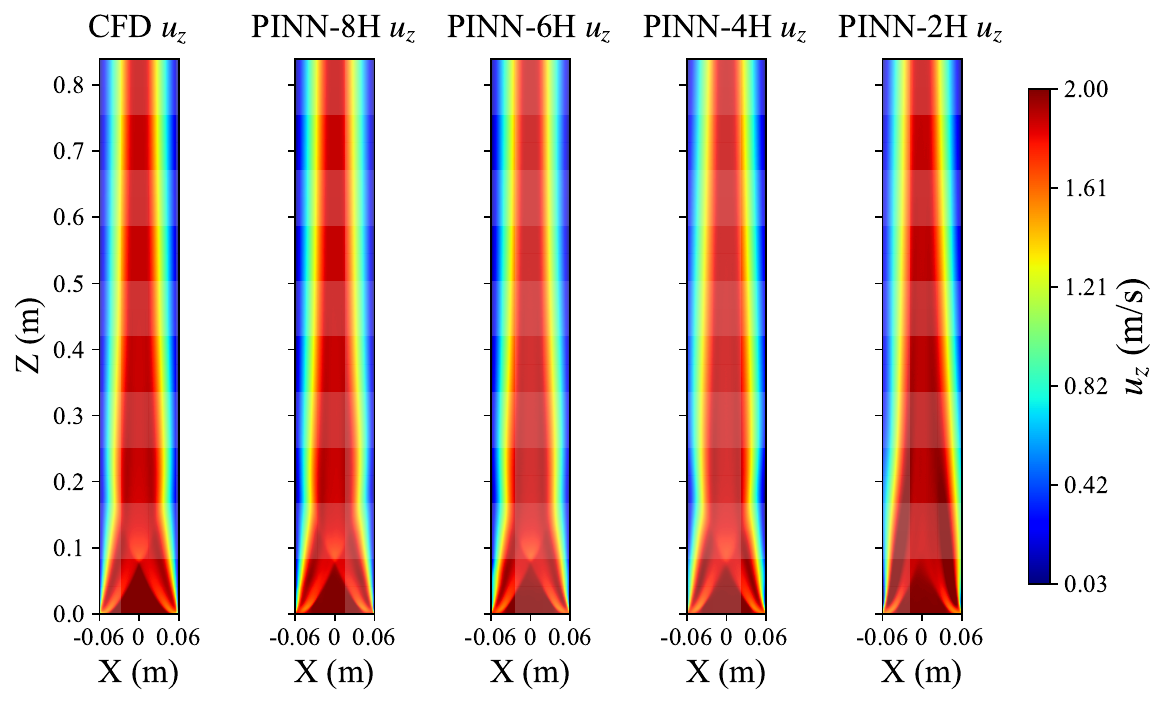}
\par(b) $u_z$
\end{minipage}

\vspace{0.3cm}

\begin{minipage}{0.48\textwidth}
\centering
\includegraphics[width=\textwidth]{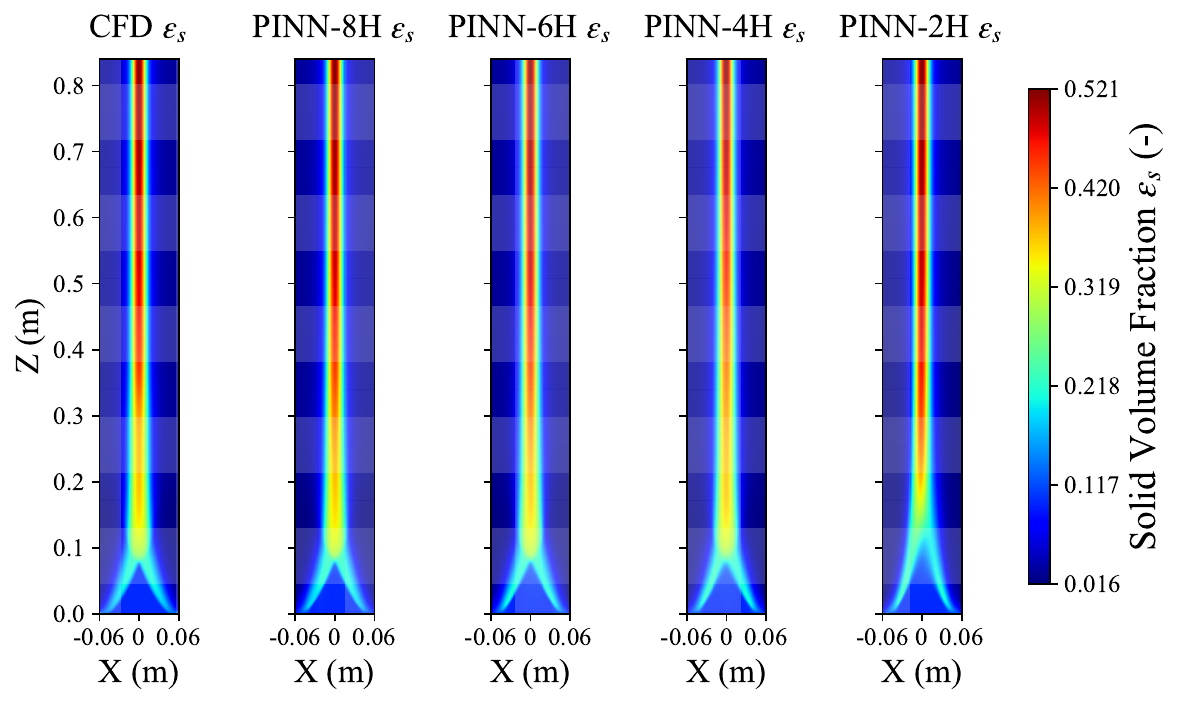}
\par(c) $\varepsilon_s$
\end{minipage}
\hfill
\begin{minipage}{0.48\textwidth}
\centering
\includegraphics[width=\textwidth]{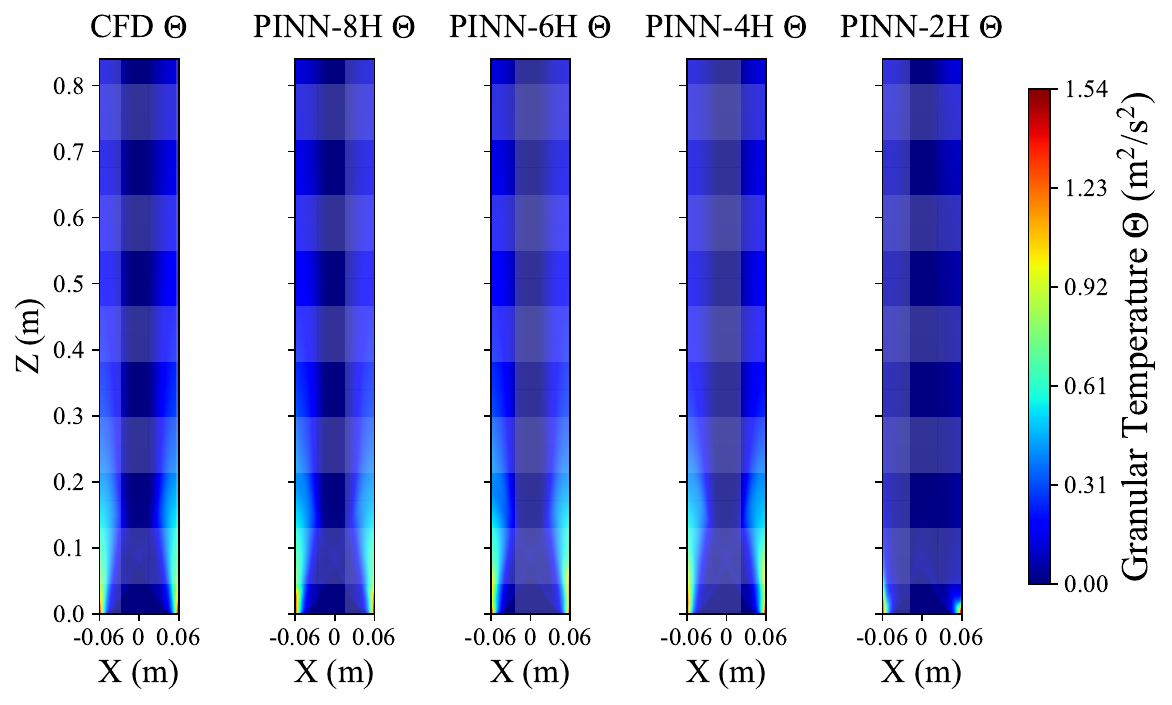}
\par(d) $\Theta_s$
\end{minipage}

\caption{Comparison between PINN-reconstructed and CFD reference flow fields for different numbers of supervision heights}
\label{pinn_cfd_comparison_Supplementary_Materials}
\end{figure}

\begin{figure}[htbp]
      \centerline{\includegraphics[width=0.48\textwidth]{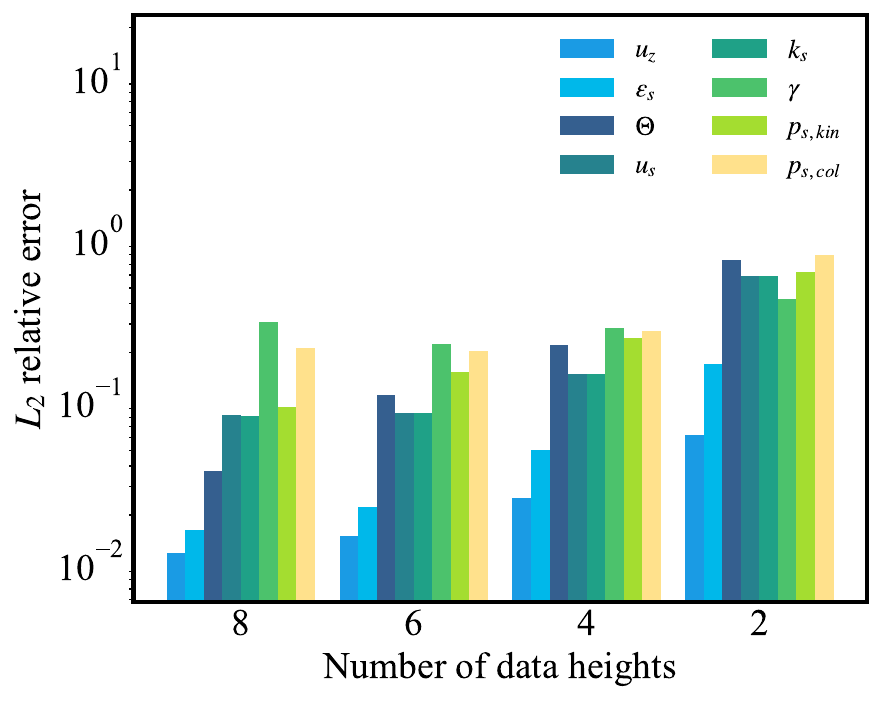}}
      \caption{Fixed-height data selection strategy} \label{16_L2_rel_all_comparison}
\end{figure}

From the flow-field contours obtained under different data-volume conditions (Figure~\ref{pinn_cfd_comparison_Supplementary_Materials}) and the corresponding full-field relative-error plots (Figure~\ref{16_L2_rel_all_comparison}), it is evident that the overall reconstruction accuracy of the PINN for granular pipe flow decreases as the number of supervision heights is reduced. Nevertheless, even with relatively sparse data, the main flow structures and multiphysics-field distribution characteristics are reasonably preserved. For the axial velocity and solid volume fraction, even when the number of supervision heights is reduced from 8 to 4, the PINN accurately captures typical features such as the central high-velocity zone, particle convergence, and inlet migration behavior. The reconstructed fields maintain good overall agreement with the CFD results, with deviations occurring only in the inlet-development section and local transition regions. This indicates that, under the constraints of the governing physical laws, the PINN retains robust reconstruction capabilities for the primary flow variables. When the number of supervision heights is further reduced to 2, although the overall flow-field morphology is maintained, the prediction errors in the inlet-development section and particle-accumulation regions increase significantly, and some local details become distorted. Excessively sparse data supervision is insufficient to constrain the development of complex flow structures. The full-field relative error of the axial velocity increases from $1.1 \times 10^{-2}$ to $1.4 \times 10^{-1}$, while that of the solid volume fraction increases from $1.6 \times 10^{-2}$ to $1.0 \times 10^{-1}$, and the granular temperature relative error increases from $3.9 \times 10^{-2}$ to $7.6 \times 10^{-1}$. Under the condition with only two supervision heights, the PINN fails to accurately reconstruct the high-granular-temperature region in the inlet-development section caused by shear.

In contrast, quantities derived from the kinetic theory, such as solids viscosity and solids thermal conductivity, are more sensitive to the amount of supervision data. As the number of supervised heights decreases, learning the high-gradient regions and local peak structures becomes progressively more difficult, particularly in the inlet and particle-aggregation regions, resulting in a significant decline in reconstruction accuracy. The variations in the full-field relative errors further show that when the number of supervision heights is reduced from 8 to 6, the increase in error for each physical quantity is relatively limited, indicating that reducing the supervision data within a certain range does not severely weaken the overall reconstruction capability. However, when the number of supervision heights is reduced to 4 or 2, the errors increase significantly, particularly for complex variables such as solid-phase pressure and collisional dissipation, for which the error growth is more pronounced. The axial velocity and solid volume fraction maintain high accuracy with 8 or 6 supervision heights, whereas the errors of transport coefficients such as the collisional pressure and kinetic pressure increase rapidly under low-data-volume conditions, mainly because of the nonlinear amplification of errors in their constitutive equations involving the solid volume fraction and granular temperature.

\begin{figure}[htbp]
\centering

\begin{minipage}{0.4\textwidth}
\centering
\includegraphics[width=\textwidth]{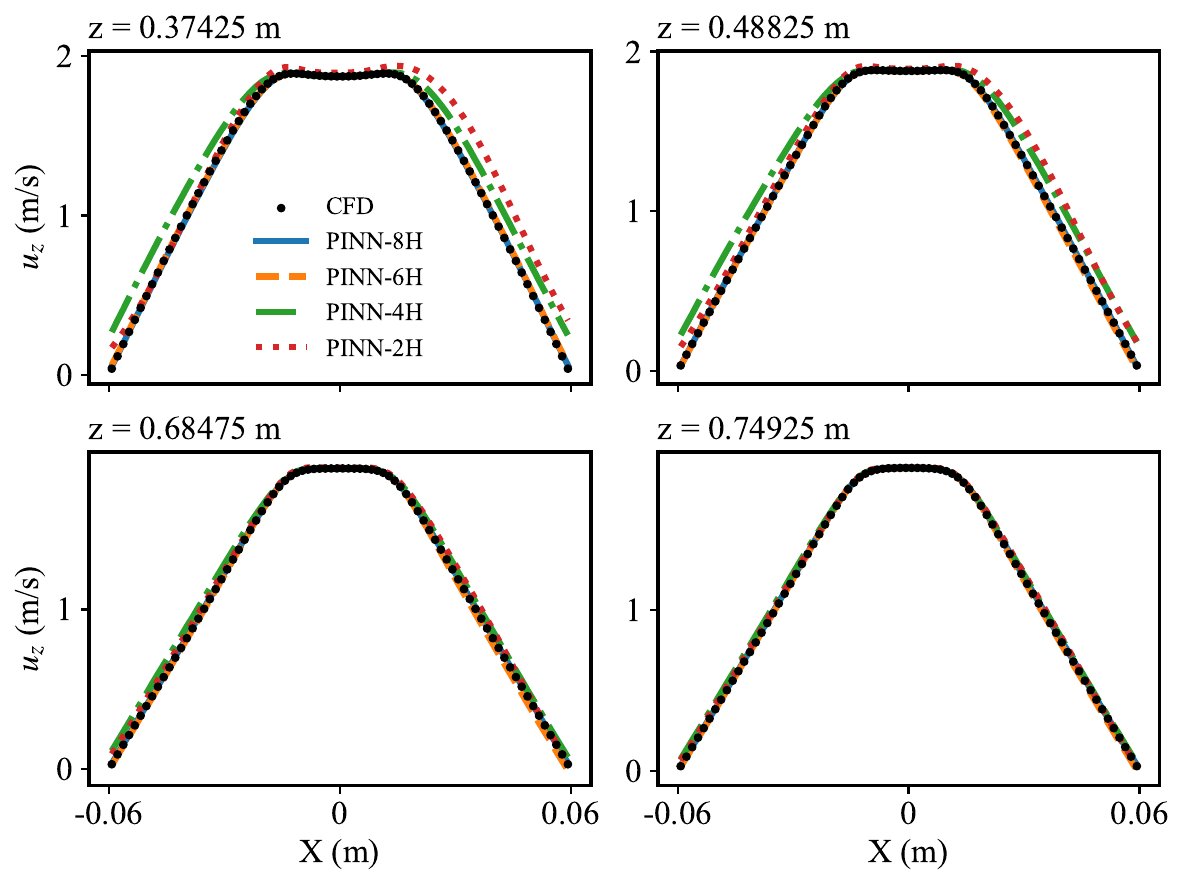}
\par(a) $u_z$
\end{minipage}
\hfill
\begin{minipage}{0.4\textwidth}
\centering
\includegraphics[width=\textwidth]{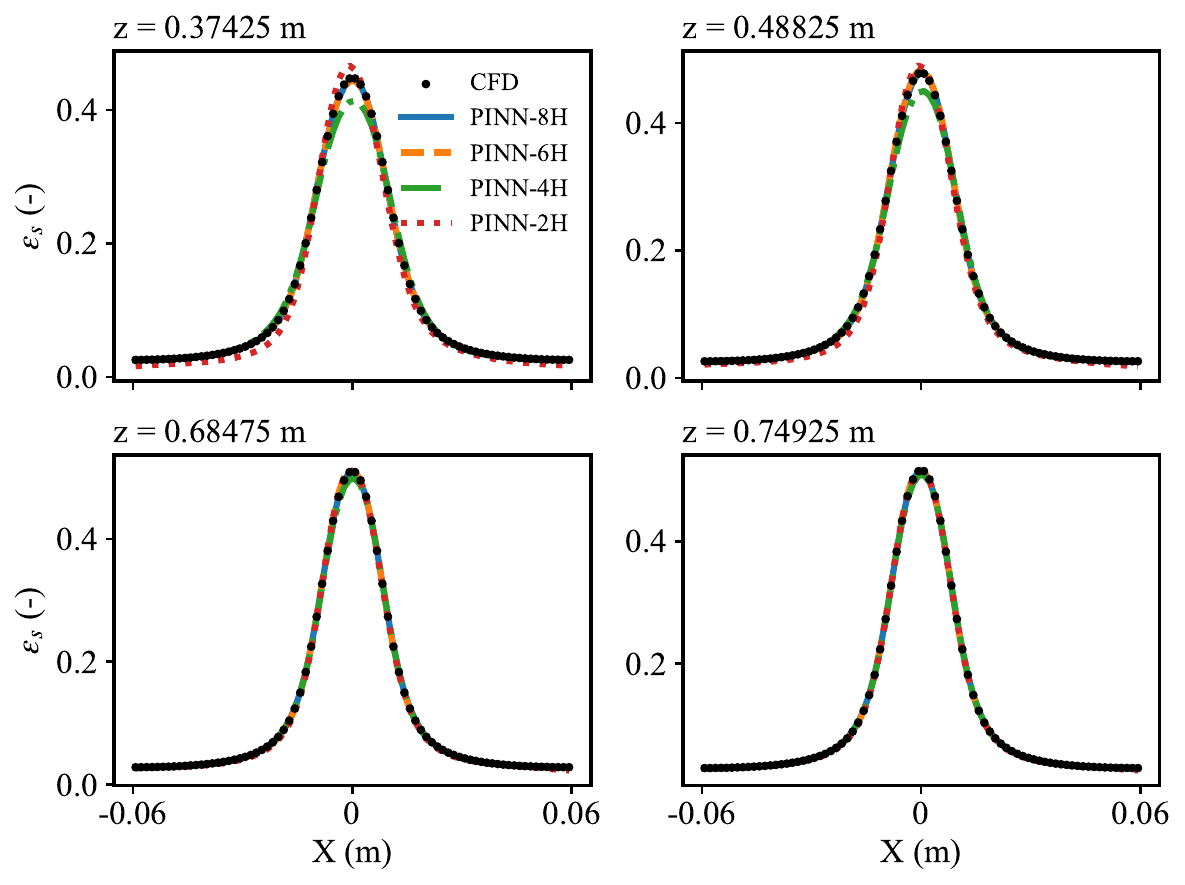}
\par(b) $\varepsilon_s$
\end{minipage}

\vspace{0.3cm}

\begin{minipage}{0.4\textwidth}
\centering
\includegraphics[width=\textwidth]{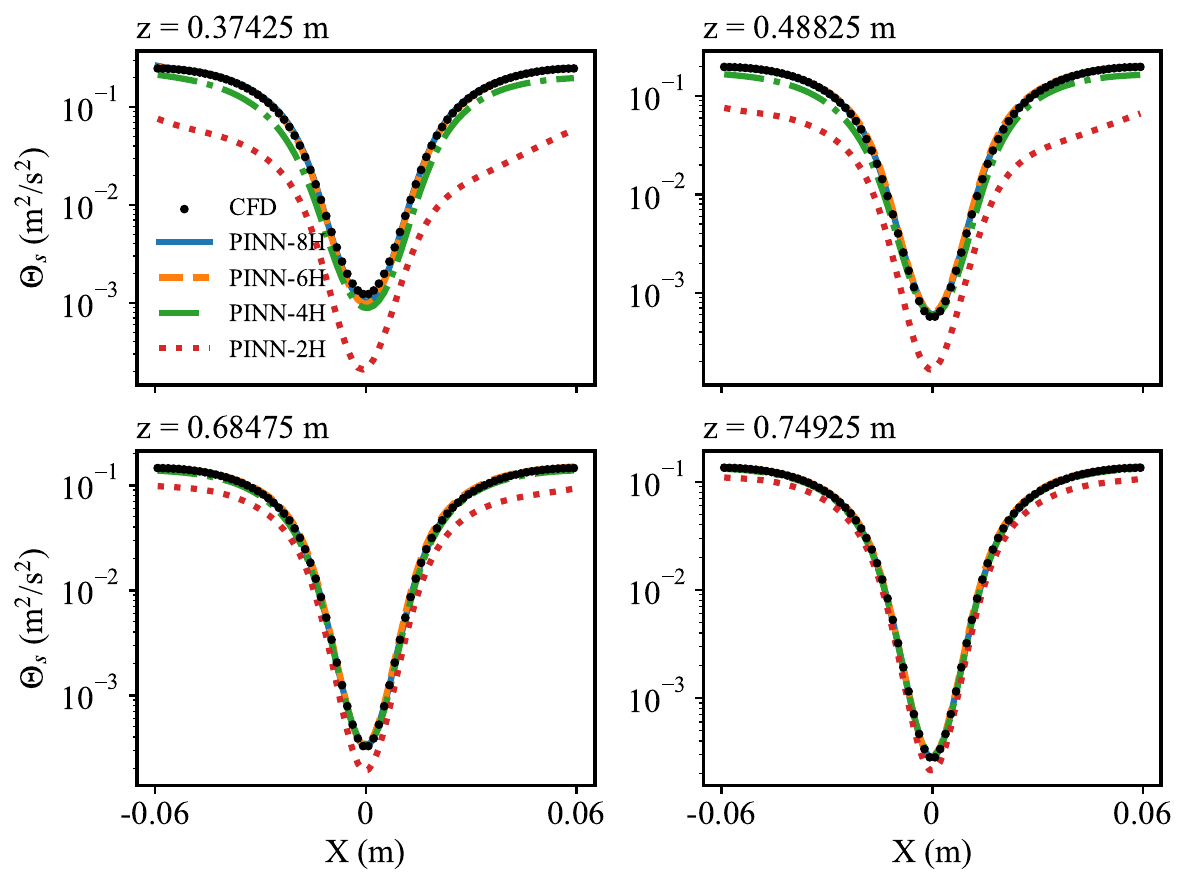}
\par(c) $\Theta_s$
\end{minipage}
\hfill
\begin{minipage}{0.4\textwidth}
\centering
\includegraphics[width=\textwidth]{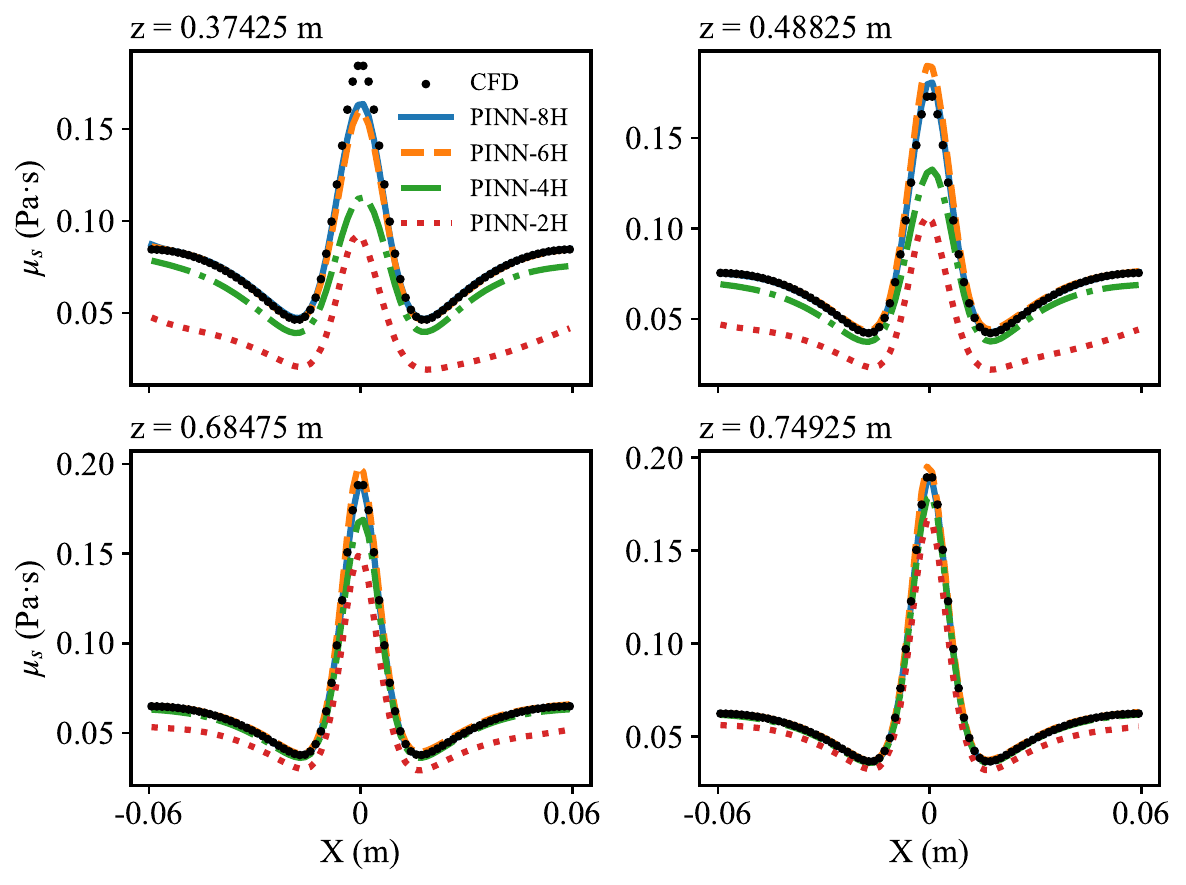}
\par(d) $\mu_s$
\end{minipage}

\vspace{0.3cm}

\begin{minipage}{0.4\textwidth}
\centering
\includegraphics[width=\textwidth]{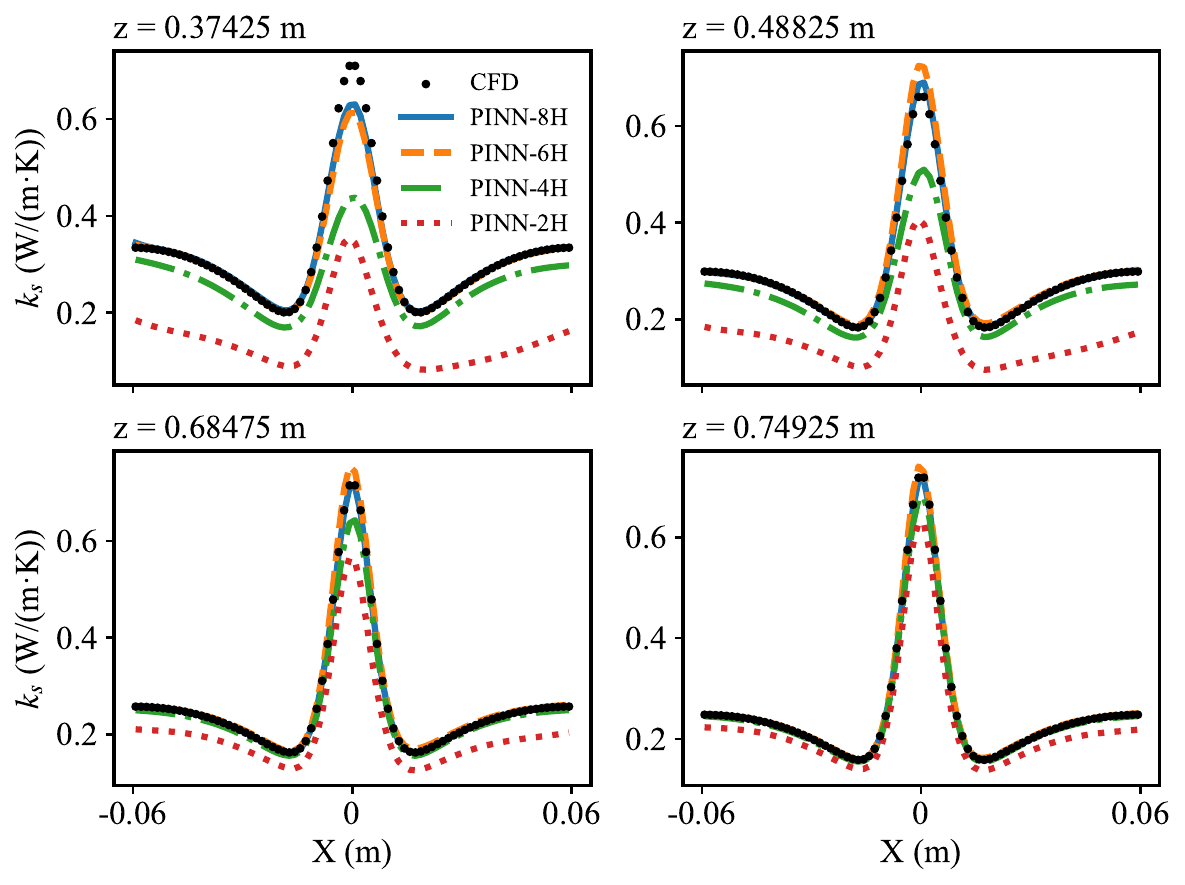}
\par(e) $k_s$
\end{minipage}
\hfill
\begin{minipage}{0.4\textwidth}
\centering
\includegraphics[width=\textwidth]{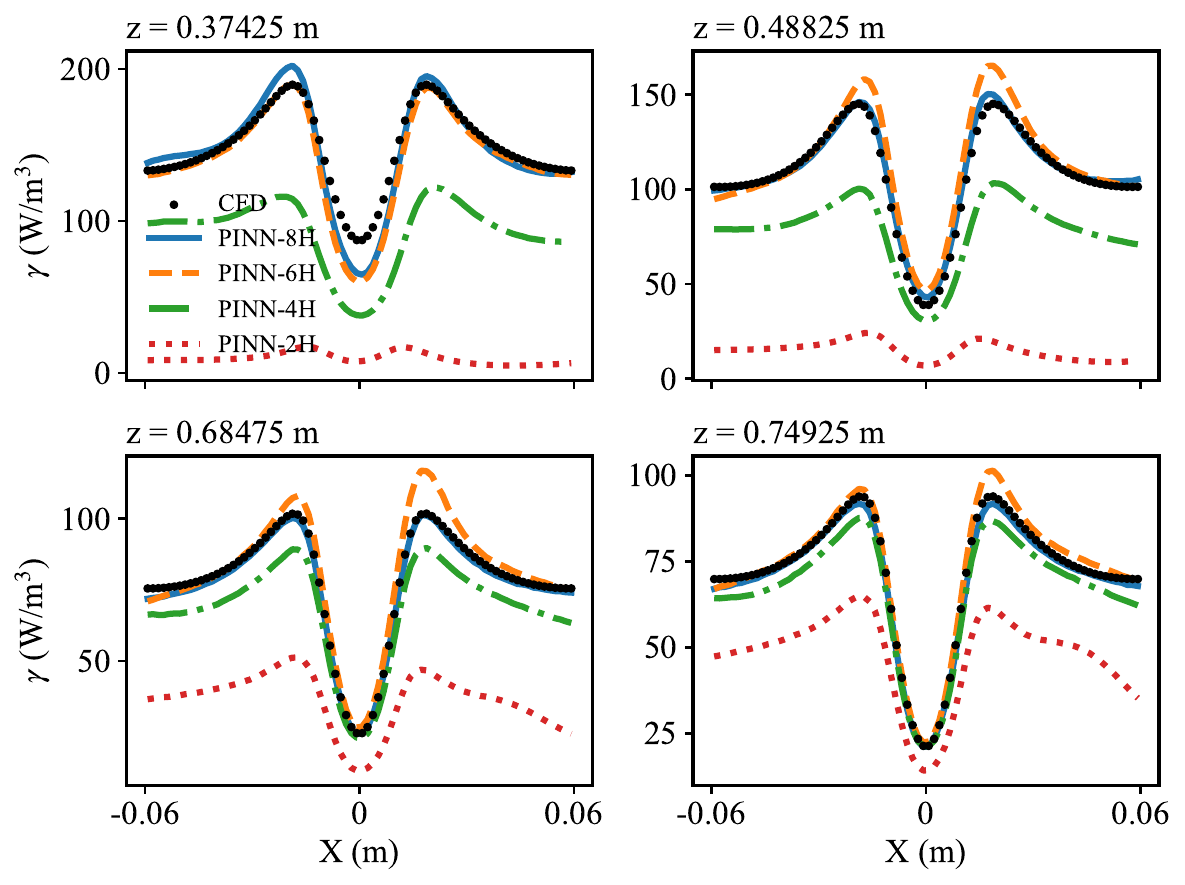}
\par(f) $\gamma$
\end{minipage}

\vspace{0.3cm}

\begin{minipage}{0.4\textwidth}
\centering
\includegraphics[width=\textwidth]{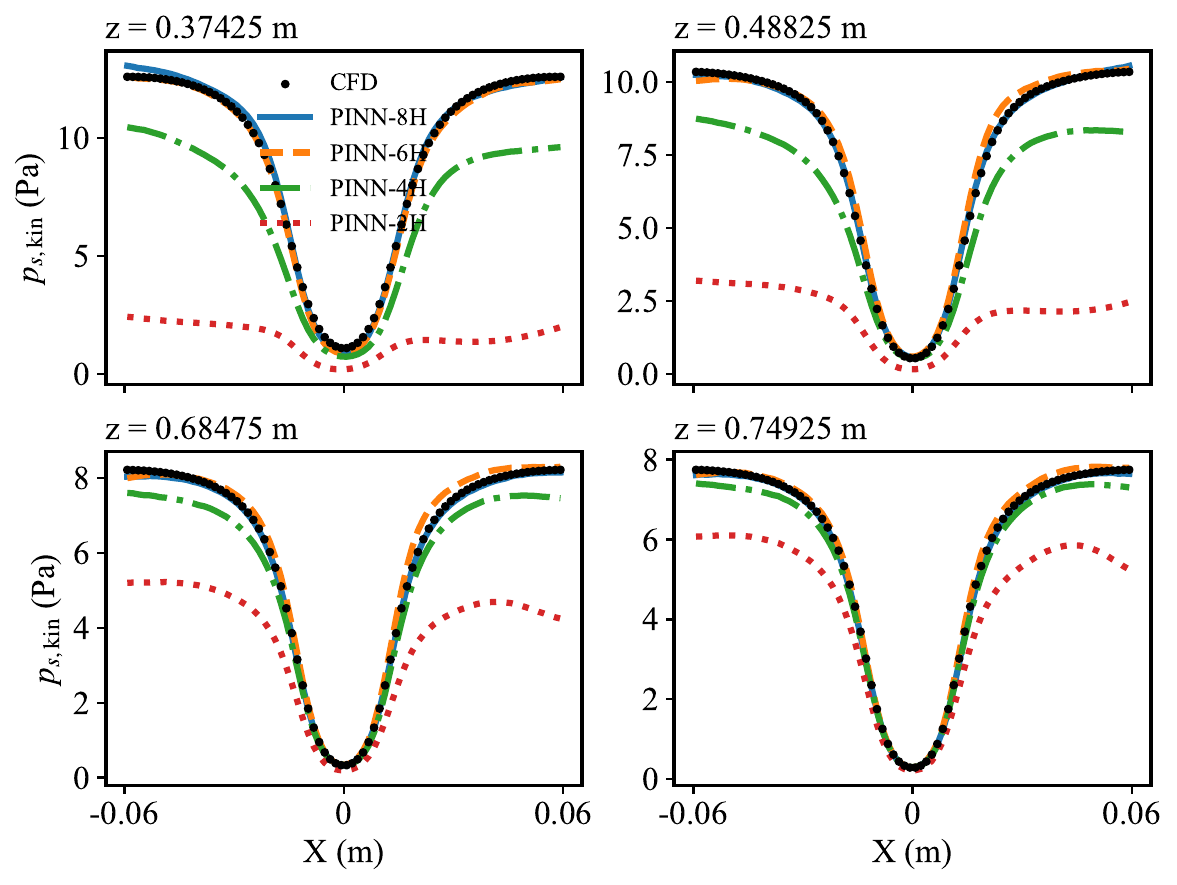}
\par(g) $p_{s,kin}$
\end{minipage}
\hfill
\begin{minipage}{0.4\textwidth}
\centering
\includegraphics[width=\textwidth]{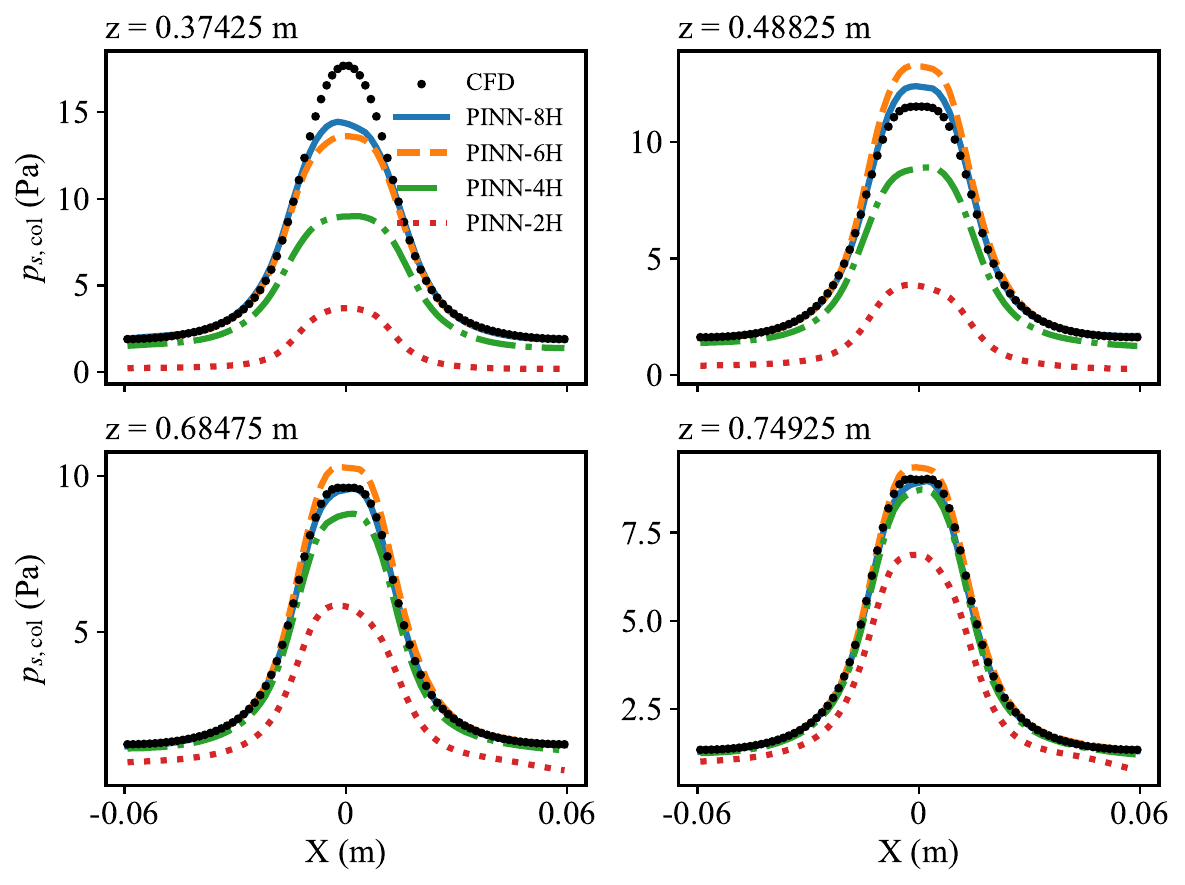}
\par(h) $p_{s,col}$
\end{minipage}

\caption{PINN predictions obtained using different numbers of supervision heights in the fully developed region ($z$ = 0.37425, 0.48825, 0.68475, and 0.74925 m), including axial velocity $u_z$, solid volume fraction $\varepsilon_s$, and granular temperature $\mathrm{\Theta}_s$. The reconstructed flow field was further used with KTGF constitutive relations to compute $\mu_s$, $k_s$, $\gamma$, $p_{s,kin}$, and $p_{s,col}$, which were compared with the corresponding CFD reference values.}
\label{PINN_predictions_Supplementary_Materials}
\end{figure}

\begin{figure}[htbp]
    \centering

    \begin{minipage}{0.46\textwidth}
        \centering
        \includegraphics[width=\textwidth]{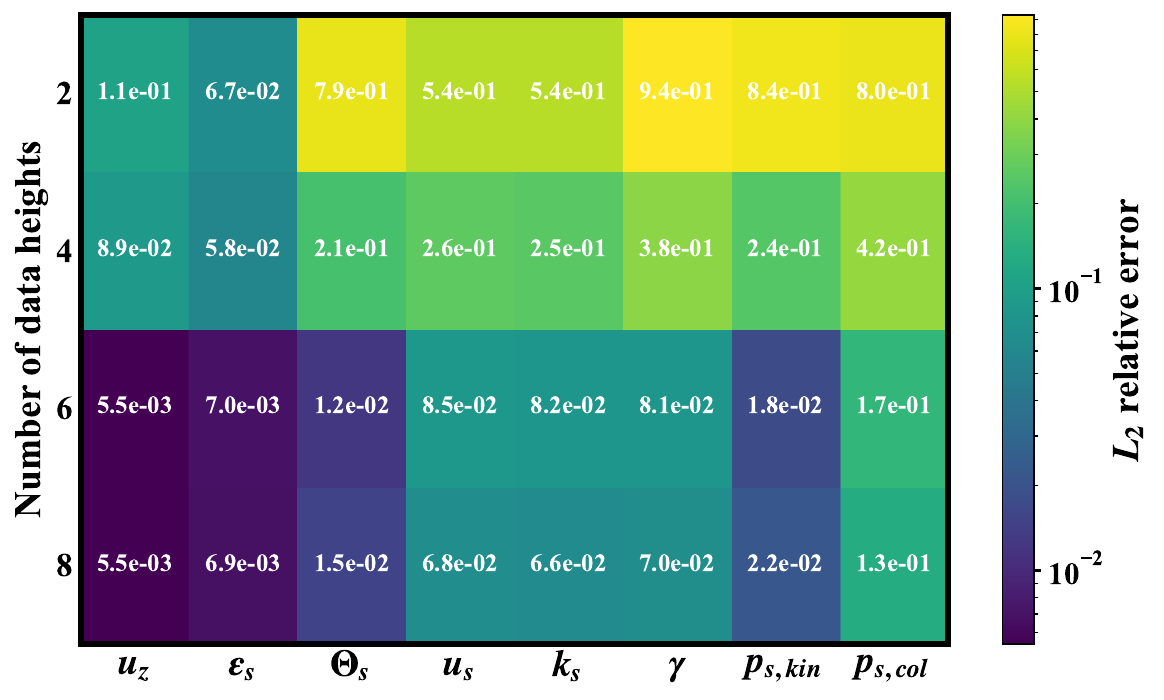}
        \par(a) $z=0.37425$ m
    \end{minipage}
    \hfill
    \begin{minipage}{0.46\textwidth}
        \centering
        \includegraphics[width=\textwidth]{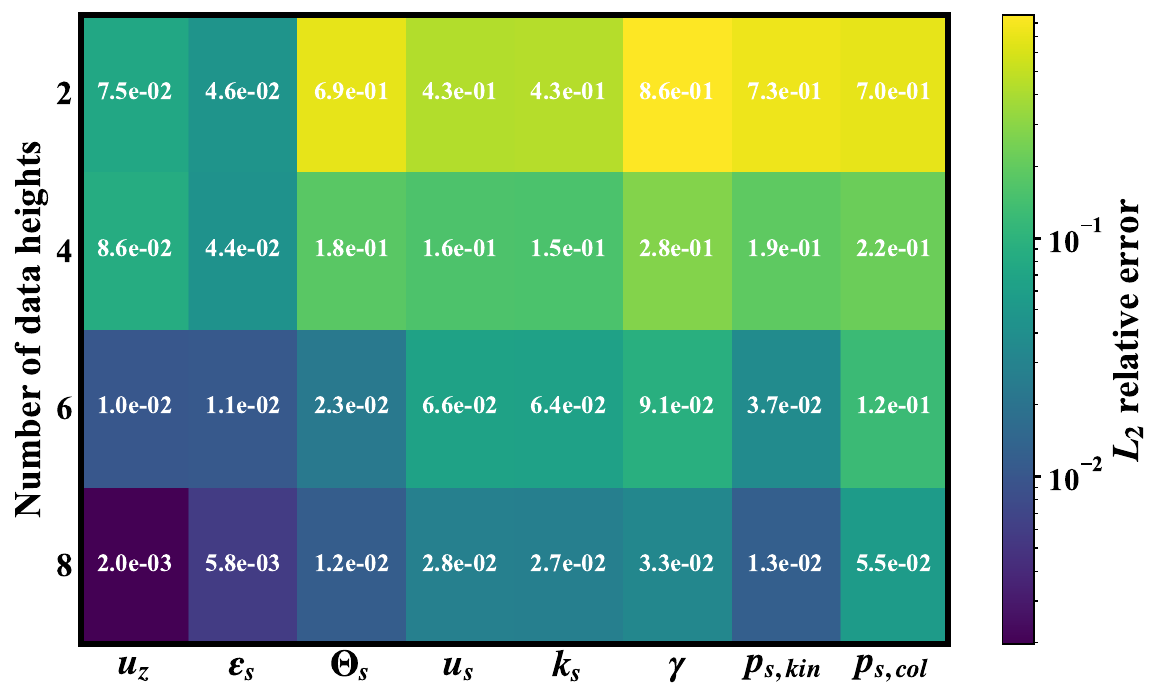}
        \par(b) $z=0.48825$ m
    \end{minipage}

    \vspace{0.3cm}

    \begin{minipage}{0.46\textwidth}
        \centering
        \includegraphics[width=\textwidth]{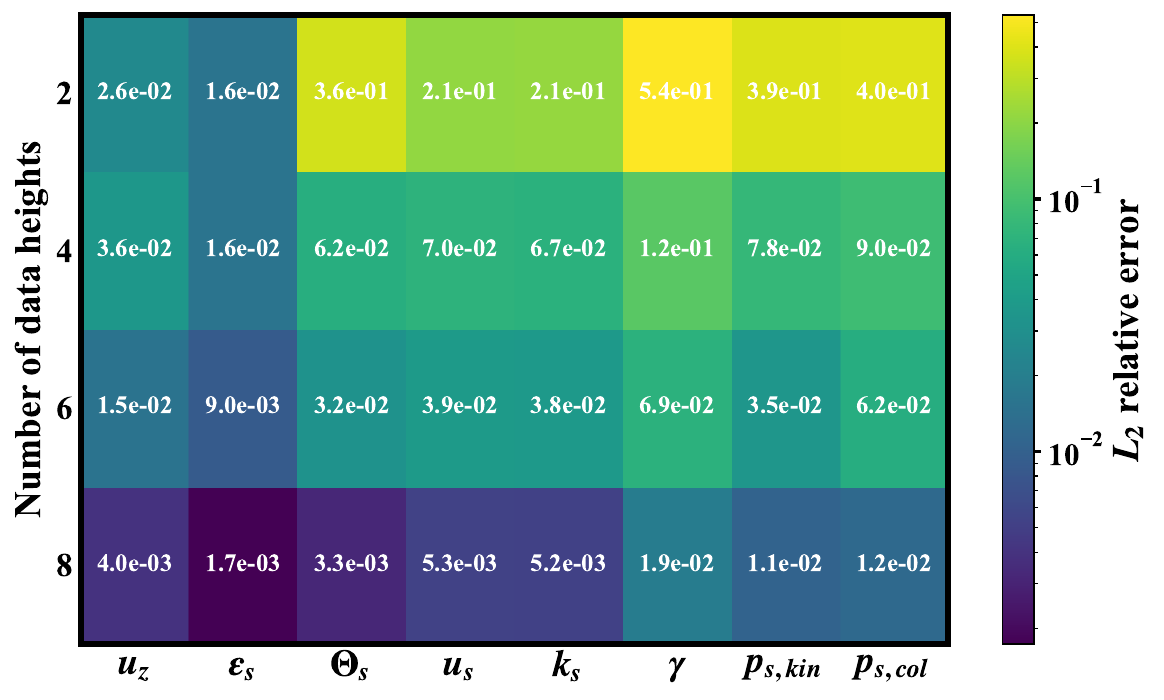}
        \par(c) $z=0.68475$ m
    \end{minipage}
    \hfill
    \begin{minipage}{0.46\textwidth}
        \centering
        \includegraphics[width=\textwidth]{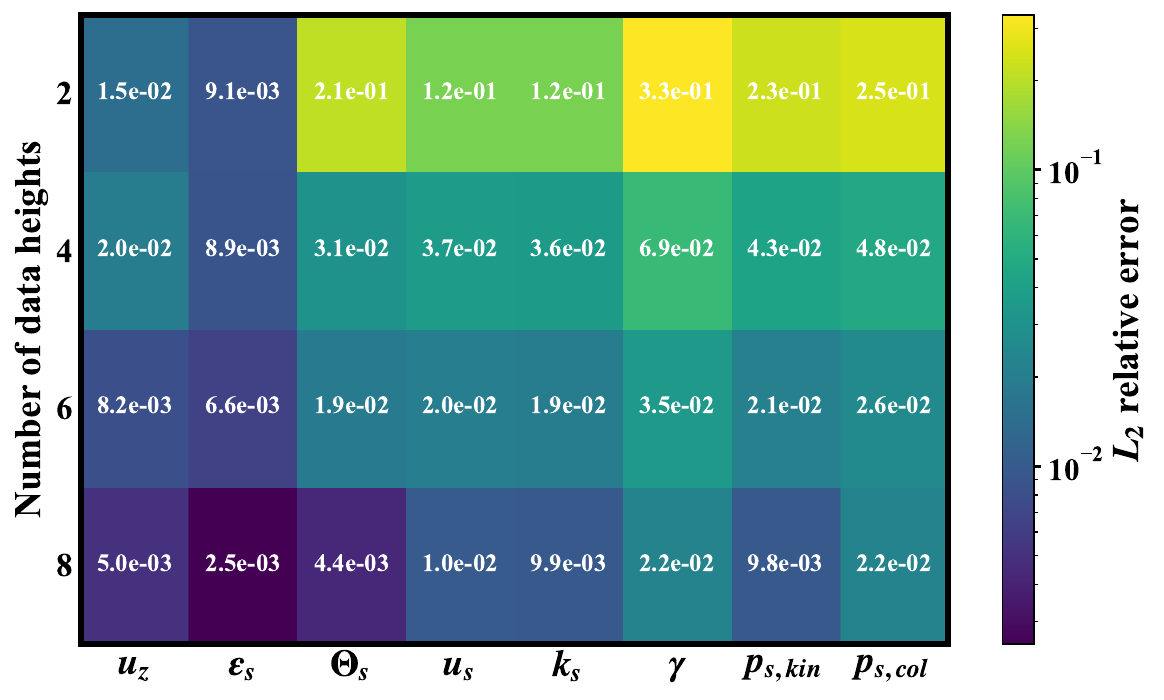}
        \par(d) $z=0.74925$ m
    \end{minipage}

    \caption{Relative errors of physical quantities at $z=0.37425$, $0.48825$, $0.68475$, and $0.74925$ m for different numbers of supervision heights.}
    \label{relative_errors_different_Supplementary_Materials}
\end{figure}

To further evaluate the PINN's reconstruction capability for local structures under different amounts of supervision data, Figure~\ref{PINN_predictions_Supplementary_Materials} presents comparisons of the radial distributions of various physical quantities at multiple axial positions. As shown in Figures~\ref{PINN_predictions_Supplementary_Materials} and~\ref{relative_errors_different_Supplementary_Materials}, as the number of supervision heights decreases from 8 to 2, the prediction errors of all variables at the corresponding analysis heights gradually increase. However, different physical quantities exhibit varying sensitivities to data sparsity. Overall, the direct network outputs---axial velocity and solid volume fraction---maintain good prediction stability even with reduced supervision data. In contrast, granular temperature and the transport coefficients derived from the kinetic theory, including solids viscosity, thermal conductivity, collisional dissipation, and solid-phase pressure, exhibit significant increases in errors at local peaks and in high-gradient regions as the amount of supervision information decreases.

For the axial velocity, the PINN accurately captures the radial distribution characteristics under all supervision conditions, with only minor deviations at the pipe center. This robustness likely stems from the strong constraint imposed by overall momentum conservation. For solids viscosity, solids thermal conductivity, and collisional dissipation, the errors increase significantly as the constraint provided by the supervision heights decreases. With only 2 supervision heights, the overall distributions deviate substantially from the CFD results, indicating that additional data constraints are required to improve the prediction accuracy. These cross-sectional distribution results further confirm that the PINN reconstruction capability for particle-laden pipe-flow structures under sparse-data conditions is strongly correlated with the amount of supervision data. When sufficient data are available, the PINN accurately reconstructs the local distribution characteristics of multiple physical variables; as the amount of supervision data decreases, the reconstruction errors accumulate rapidly, particularly in the inlet-development region.

In summary, the PINN can reconstruct the complete flow-field information from sparse data, but its accuracy is closely related to the amount of supervision data. For the direct network outputs, such as axial velocity and solid volume fraction, the PINN maintains relatively high accuracy even when the amount of data is reduced. However, for granular temperature and the KTGF-derived transport coefficients, the prediction accuracy decreases significantly as the amount of data is reduced, resulting in partial distortion of the flow field. This does not indicate model failure, particularly considering that the supervision data points may account for only 1.4\% of the complete flow-field information. In practical applications, a balance must be achieved between measurement cost and reconstruction accuracy, and the number and locations of the supervision sections should be carefully arranged to enhance the reliability of the PINN in solving complex particle-laden multiphase flows.

Compared with the fixed-height supervision method, the random sampling strategy provides broader spatial coverage and slightly higher overall PINN reconstruction accuracy. However, fixed-height data supervision is more representative of actual experimental measurement conditions. The results demonstrate that, under limited data supervision, the PINN can still accurately reconstruct the flow-field distributions in granular pipe flow, with higher accuracy in the fully developed region, indicating good engineering application potential for the PINN method based on fixed-section measurement data.

The amount of supervision data significantly affects the PINN reconstruction accuracy. As the number of supervision heights gradually decreases, the overall errors of all physical quantities tend to increase. Among these quantities, the axial velocity and solid volume fraction exhibit strong robustness to data sparsification, whereas the granular temperature, solids thermal conductivity, solids viscosity, collisional dissipation, and collisional pressure are more sensitive to the amount of supervision data. When the supervision data become excessively sparse, local distortion occurs in the inlet region of the flow field. In practical engineering applications, measurement cost and prediction accuracy must be considered comprehensively. By rationally arranging the positions of the supervision sections, the generalization capability and engineering applicability of the PINN for complex particle-laden flows can be improved.

\section*{Reference}

\bibliographystyle{elsarticle-harv}
\bibliography{Reference}
\end{spacing}
\end{document}